\documentclass[a4paper,12pt]{article}
\usepackage[english]{babel}
\usepackage[utf8]{inputenc}
\usepackage[T1]{fontenc}
\usepackage{lmodern}
\usepackage{amsmath,amssymb,amsfonts,bm}
\usepackage{graphicx,booktabs,longtable,array,tabularx}
\usepackage{geometry,authblk,placeins,caption,setspace,microtype}
\usepackage[authoryear]{natbib}
\setcitestyle{authoryear,open={(},close={)}}
\usepackage[hypertexnames=false,hidelinks]{hyperref}
\graphicspath{{figures/}}
\newcommand{\notetext}[1]{\par\vspace{4pt}{\footnotesize\raggedright #1\par}}
\newcommand{\figone}[4]{\begin{figure}[!htbp]\centering\includegraphics[width=#4\textwidth]{#1}\caption{#2}\label{#3}\end{figure}}

\newcommand{\Nobs}{1990}
\newcommand{\Nedges}{214}
\newcommand{\Nmaterial}{312}
\newcommand{\Nbackbone}{32}

\newcommand{\BootSIS}{0.50}
\newcommand{\BootSVS}{0.72}
\newcommand{\SISsize}{0.30}
\newcommand{\SVSsize}{0.46}
\newcommand{\TopFiveSIS}{51.4}
\newcommand{\TopFiveSVS}{28.2}
\newcommand{\MatchedSISRho}{0.63}
\newcommand{\MatchedSISTopFive}{3}
\newcommand{\MatchedSISMedianRankShift}{7.0}
\newcommand{\MatchedSVSRho}{0.66}
\newcommand{\MatchedSVSTopFive}{3}
\newcommand{\MatchedSVSMedianRankShift}{5.0}

\title{\textbf{Contemporaneous Transmission Networks and Systemic Risk in European Banking}}
\ifdefined\anonymousversion
\author{}\date{}
\else
\author[1,2]{Minh Duc Duong}
\author[3]{Alessio Moneta}
\author[4]{Fabrizio Lillo}
\author[1]{Caterina Giannetti}
\author[5]{Tu Anh Doan Nguyen}
\affil[1]{University of Pisa, Pisa, Italy}
\affil[2]{IMT School for Advanced Studies Lucca, Lucca, Italy}
\affil[3]{Sant'Anna School of Advanced Studies, Pisa, Italy}
\affil[4]{Scuola Normale Superiore, Pisa, Italy}
\affil[5]{University of Verona, Verona, Italy}
\date{October 2026}
\fi
\begin{document}
\maketitle
\begin{abstract}
We study how contemporaneous equity-return transmission shapes systemic exposure among 51 listed European banks. Using a recursive non-Gaussian network model on volatility-standardized returns, we construct two market-capitalization-weighted measures based on direct and indirect transmission effects: the Systemic Impact Score (SIS), capturing a bank’s potential influence on other institutions, and the Systemic Vulnerability Score (SVS), capturing its exposure to disturbances originating elsewhere. The fitted network reveals distinct transmitting and receiving roles, highlighting differences between institutions with substantial outward influence and those with high incoming sensitivity. Receiving roles are generally more reproducible than transmitting roles across repeated network discovery, and selected connections remain informative about lower-tail dependence in subsequent observations. Volatility standardization substantially reduces detected dependence among recovered shocks, although the full-sample model does not satisfy the designated criterion for supporting shock independence. Alternative recursive orderings also yield comparable directional tail asymmetries, limiting the extent to which these patterns validate a particular ordering. The results provide complementary measures of model-implied systemic transmission and vulnerability, with explicit attention to uncertainty and identification limits.
\end{abstract}
\noindent\textbf{Keywords:} systemic risk; European banks; financial networks; LiNGAM; contemporaneous transmission.\\
\textbf{JEL:} C32, C58, G01, G21, G28.
\clearpage

\section{Introduction}

Understanding how financial shocks spread across banks is central to the study of systemic risk. Although large institutions are generally considered systemically important, size alone does not reveal how distress moves through the banking system, as systemic relevance also depends on the position of an institution  within the network through which shocks propagate \citep{battiston2012debtrank,glasserman2016contagion}. A bank may originate shocks, amplify shocks received from elsewhere, or primarily absorb disturbances transmitted by other institutions. Theoretical models further show that interconnectedness is not unambiguously stabilizing or destabilizing: interbank links can facilitate risk sharing and liquidity redistribution, while their effects on systemic fragility depend on network structure and on whether shocks are idiosyncratic or correlated \citep{castiglionesi2018liquidity,castiglionesi2019financial}. Therefore, distinguishing between institutions that predominantly transmit shocks and those that predominantly absorb them is  important for macroprudential policy because institutions of similar size may occupy very different positions in the transmission process.  
Existing market-based measures, such as $\Delta$CoVaR, MES, and SRISK, provide timely indicators of systemic contribution and vulnerability but generally do not identify the pathways through which shocks propagate \citep{benoit2017risks,brownlees2017srisk,acharya2025systemic}. By contrast, models that explicitly represent contagion through interbank exposures, liquidity channels, or asset fire sales, provide a more detailed account of these pathways but often require granular or confidential data and are therefore difficult to implement on a broad and regular basis \citep{montagna2020origin,halaj2025decomposing}.

The approach we present in this paper is complementary to these structural models: using publicly available market data, we estimate  a directed contemporaneous transmission network among European banks and construct measures of systemic impact and vulnerability from the estimated network. We employ the VAR-LiNGAM (Vector Autoregressive - Linear Non-Gaussian Model) framework, which combines a VAR model of lagged relationships with a causal-discovery approach to identify the direction and magnitude of contemporaneous effects \citep{Shimizu06a,hyvarinen2010}. Building on the broader literature on directed acyclic graphs and causal discovery \citep{pearl2000causality,Peters2017,glymour2019review}, VAR-LiNGAM and related methods have been applied in several economic and financial settings, including firm growth \citep{Moneta_2013,ciarli2023does}, monetary policy \citep{moneta2004identification}, and international financial markets \citep{hyvarinen2010,jiang2023}. In our application, we build on this framework by using the estimated contemporaneous network to derive a total-effect matrix that captures both direct effects and effects transmitted indirectly through other institutions. From this matrix, we construct two market-capitalization-weighted measures. The \emph{Systemic Impact Score} (SIS) measures the extent to which a shock originating from a bank is transmitted to the rest of the system, while the \emph{Systemic Vulnerability Score} (SVS) measures the extent to which a bank is affected by shocks originating from other institutions. The central idea is to use the joint behavior of bank returns to distinguish institutions that act mainly as transmitters from those that act mainly as receivers. This distinction is particularly relevant in financial markets, where shocks may spread rapidly across institutions. With daily data, the originating shock and its effects may be recorded within the same trading day rather than appear as return movements that predict those of subsequent days.

This interpretation, however, relies on specific identifying assumptions. Contemporaneous relationships are assumed to be linear and acyclic, implying a recursive ordering of transmission within each observation period, while the underlying structural shocks must be non-Gaussian and mutually independent. With daily data, this recursiveness should be interpreted as an approximation to same-day transmission: interactions occurring within the trading day are compressed into the same observation, so intraday feedback cannot be separately identified. The model additionally requires causal sufficiency; including a broad market proxy cannot establish that all relevant sovereign, country or sector common causes are observed. Under these assumptions, the contemporaneous ordering and associated transmission coefficients can be identified using LiNGAM-type methods. We implement the causal-discovery component using DirectLiNGAM \citep{shimizu2011directlingam}. In a general VAR-LiNGAM specification, a VAR first accounts for linear lagged dependence, and DirectLiNGAM is then applied to the resulting innovations. In our application, the lag-selection analysis favors $p=0$, meaning that no lagged terms are included in the baseline model. DirectLiNGAM is therefore applied to mean-centered and volatility-standardized daily returns; the unfiltered-return specification remains a sensitivity. A VAR$(1)$-LiNGAM specification is retained as a robustness check to examine whether accounting for one-day predictive relationships changes the recovered network. The VAR$(1)$ comparison measures sensitivity to removing one retained-observation lag; source rankings do change to a meaningful degree, so this exercise is not treated as evidence of exact structural invariance.

Relative to existing approaches to systemic-risk measurement, our contribution is to estimate model-implied directional contemporaneous transmission from publicly available market data and evaluate the assumptions underlying its possible structural interpretation. Econometric connectedness measures based on principal components, Granger-causality networks, or forecast-error variance decompositions capture common variation, predictive relationships, and directional spillovers \citep{billio2012econometric,diebold2014network}, while network approaches based on quantile regressions, tail dependence, and centrality identify risk spreaders and absorbers \citep{billio2017systemic}. These approaches provide informative measures of interconnectedness without imposing the particular non-Gaussian independent-noise assumptions used to orient same-period coefficients here. At the other end of the spectrum, structural contagion models based on interbank exposures, liquidity channels, or common asset holdings can identify specific mechanisms but typically require granular, and often confidential, information \citep{halaj2025decomposing}. The key distinction is therefore not the use of market data per se, but the conditional structural interpretation assigned to same-period transmission. This interpretation relies on stronger identifying assumptions, whose empirical plausibility we examine through independence, recursiveness, and bootstrap diagnostics, kept as distinct and non-compensatory pieces of evidence. Our empirical assessment draws on the post-recovery evidentiary framework of \citet{doannguyenbucciol2026}, which distinguishes the recovery of a contemporaneous representation from empirical support for the identifying independence assumption and the reproducibility of recovered relations. The two latter forms of evidence are not logically interchangeable, as established under the constructions and assumptions of their Proposition A.1. We therefore assess recovered-shock dependence and discovery uncertainty separately, while retaining SIS and SVS as the principal economic outputs of the banking analysis.

A smaller literature has applied explicit causal-discovery methods to financial stress and transmission. \cite{GAO2018} develop a stress-testing framework based on Suppes--Bayes causal networks, while \cite{RIGANA2023} use a causal-network approach to identify contagion pathways in foreign-exchange markets. Related work uses causal-search methods to study cross-market linkages, including foreign-exchange, stock and bond markets \citep{jiang2023}. \cite{kumar2023} provide a broader survey of causal-inference applications in banking, finance, and insurance. Relative to this literature, we apply causal discovery to a broad cross-section of European banks and use the recovered total-effect structure to construct institution-level measures of systemic impact and vulnerability. We further examine how uncertainty in structural discovery propagates from individual edge directions to the resulting systemic-risk rankings, allowing us to distinguish uncertainty about particular channels from the reproducibility of broader transmitting and receiving roles. These assessments qualify the interpretation of the source--receiver distinction rather than replacing it as the empirical contribution.

Our empirical analysis addresses three related questions. Which banks emerge as important sources of contemporaneous transmission, and which are more exposed to disturbances originating elsewhere in the banking network? How sensitive are these roles to volatility treatment and uncertainty in the recovery of network directions? And do the estimated connections retain economic relevance for downside co-movements outside the estimation period?

By considering these questions jointly, we distinguish the economic roles described by SIS and SVS from the strength of the evidence supporting their structural interpretation. This allows us to assess systemic impact and vulnerability while explicitly acknowledging uncertainty in network discovery and the identifying assumptions underlying the estimated transmission channels.

The remainder of the paper is organized as follows. Section 2 reviews the literature on systemic-risk measurement, financial connectedness, and network contagion, and positions our contribution relative to both reduced-form and structural approaches. Section 3 presents the VAR-LiNGAM framework, discusses the identifying assumptions underlying contemporaneous transmission, defines the SIS and SVS measures, and introduces the procedures used to assess coefficient significance, directional stability, recursiveness, and shock independence. Section 4 describes the sample of European banks, the market data, and the construction of the variables used in the analysis. Section 5 reports the estimated transmission network and systemic-risk rankings, examines their robustness to structural uncertainty and alternative lag specifications, evaluates the economic relevance of the estimated directions using tail diagnostics, and compares SIS and SVS with Diebold--Yilmaz connectedness and MES. Section 6 concludes by summarizing the main economic implications, discussing the limits of the contemporaneous causal interpretation, and relating the empirical findings to theoretical work on financial networks.

\section{Literature Review}

Measuring systemic risk has become a cornerstone of both academic research and financial regulation, with an expanding set of methodologies seeking to capture how fragility builds up and propagates across institutions. \cite{benoit2017risks} distinguish between measures based on specific contagion and amplification mechanisms and global market-based measures, such as $\Delta$CoVaR, MES, and SRISK. The former are closely linked to theoretical models and can identify particular channels of interconnectedness, but are often model-dependent, data-intensive, and limited to selected transmission mechanisms. The latter use market data to measure institutions' exposure or contribution to system-wide losses under stress and are therefore easier to implement and compare. However, they may largely capture conventional market risk rather than systemic externalities. Integrating the two approaches could therefore combine broader measurement with a clearer identification of the mechanisms generating systemic risk.
 \footnote{Representative market-based systemic risk measures include:
(i) $\Delta$CoVaR, which captures how much a firm contributes to system-wide tail risk;
(ii) $e\Delta$CoVaR, which measures a firm’s vulnerability to aggregate shocks;
(iii) MES (Marginal Expected Shortfall), the expected loss of a firm conditional on a market downturn;
(iv) SRISK, the expected capital shortfall of a firm in a severe market decline; and
(v) Market Leverage, a balance-sheet indicator feeding into SRISK. See \cite{acharya2025systemic} for formal definitions.}

Building on this distinction, \cite{acharya2025systemic} provide a comprehensive long-term evaluation of these measures across more than a century of U.S. financial crises, from the Panic of 1907 to the Banking Stress of 2023. Their analysis shows that $\Delta$CoVaR, $e\Delta$CoVaR, MES, and SRISK consistently predict the cross-sectional vulnerability of institutions during stress episodes, anticipating volatility, returns, balance-sheet deterioration, and even bank failures. Importantly, the measures are more informative in severe crises and complement, rather than substitute, traditional accounting indicators, offering forward-looking insights less prone to regulatory manipulation.  However, while these market-based approaches capture fragility and exposure, they remain largely silent on the causal structure of contagion, that is, how shocks originate and propagate contemporaneously across institutions. These measures have also entered the regulatory discussion on how to designate and supervise systemically important financial institutions (SIFIs). In this context, they act not as precise thresholds, but as ranking devices: they order institutions according to their expected contribution to systemic distress, thereby informing the allocation of capital surcharges and supervisory intensity.  Consistent with this, \cite{billio2017systemic} show that market-based measures such as $\Delta$CoVaR and MES are not perfectly aligned with network-based indicators of systemic importance, indicating that complementary approaches are necessary. 

Thus, a complementary line of research has turned to network-based models of systemic risk, which explicitly capture contagion mechanisms. For example, \citet{kanno2015assessing} uses a network model of how interbank defaults cascade over time, showing that Global Systemically Important Banks (G-SIBs) sit at the center of the network and that their failure could trigger widespread contagion. In a similar vein, \cite{montagna2020origin} develop a microstructural contagion model using granular data covering almost the entire euro area banking system. They simulate correlated shocks and propagate them through solvency, liquidity, and market contagion channels, finding that correlated shocks and market contagion are the main drivers of systemic risk, while solvency and liquidity channels play a more limited role. More recently, \cite{halaj2025decomposing} decompose systemic risk into contagion, common exposures, and idiosyncratic risk using supervisory balance-sheet data, providing a structural benchmark for disentangling the origins of fragility. Conceptual surveys such as \cite{jackson2025optimal} highlight the policy relevance of network centrality, showing how systemic importance can be linked to optimal regulation and capital surcharges. At the same time, as noted by \cite{silva2017analysis}, causal modeling and simulation-based approaches remain relatively underrepresented in the systemic risk literature, leaving open important questions on the mechanisms of contagion.

A related literature studies contagion through joint extreme market movements rather than through an explicitly directed network, using co-exceedance and related tail-dependence measures to capture the clustering and spillover of extreme returns \citep{bae2003coexceedance,baur2005coexceedances,chiu2015industry}. In our analysis, these measures are not used to identify the network. Instead, they provide a complementary diagnostic of its economic relevance. The network links and directions are recovered from the full distribution of returns using LiNGAM, while the tail analysis examines whether these links and directions are also reflected in downside dependence.

A parallel empirical strand measures interconnectedness through reduced-form time-series models rather than balance-sheet networks. The connectedness framework of \cite{diebold2014network}, built on generalized forecast-error-variance decompositions \citep{pesaran1998generalized}, summarizes a system by the shares of each variable's forecast-error variance attributable to shocks in the others, delivering directional ``to'' and ``from'' spillover measures and a system-wide connectedness index. In a closely related spirit, \cite{billio2012econometric} combine principal-components analysis and Granger-causality networks to measure connectedness among banks, insurers, hedge funds and brokers, documenting that connectedness rises sharply before and during crises and that the financial sector became increasingly interconnected in the run-up to 2008. These measures are attractive because they can be estimated from returns or volatility and yield an intuitive directional decomposition of spillovers. They are, however, reduced-form: they quantify how much of a bank's forecast-error variance is associated with others' innovations, which captures dynamic predictive spillover rather than the contemporaneous causal structure of transmission, and the ordering-invariant generalized decomposition does not identify the direction of same-period effects. This distinction motivates our conditional structural analysis and our use of the Diebold--Yilmaz ``to'' and ``from'' indices as order-invariant reduced-form benchmarks (Section~\ref{sec:res_benchmarks}). Similarity or disagreement between these estimands is descriptive and does not validate structural identification.

A closely related strand measures systemic relevance through tail-risk networks. \citet{hautsch2015financial} construct realized systemic risk betas linking institution-level tail risk to system tail risk while accounting for network interdependence. \citet{hautsch2014forecasting} extend this approach to forecasting in European financial networks, and \citet{betz2016systemic} study time-varying spillovers in a European bank--sovereign system. Our contribution is not the introduction of source and receiver roles themselves. Rather, SIS and SVS aggregate total contemporaneous sensitivities in a recursive return model, and we assess how re-estimating the ordering and selected coefficients affects each role's ranking. These estimands complement, rather than replace, measures of contributions to system tail risk.

At the theoretical level, systemic risk has been studied through models of contagion and network fragility. Early contributions highlight how interbank links can absorb or amplify shocks depending on their structure \citep{allen2000financial,acemoglu2015systemic,elliott2014financial}.
\cite{krause2012interbank} further distinguish the initiation from the propagation of contagion, showing that the size of the initially failing bank is important for whether contagion occurs, whereas the structure of interbank linkages is more important for the extent of its spread. Importantly, centralization is not intrinsically destabilizing. \cite{castiglionesi2018liquidity} show that a star-shaped interbank network can efficiently channel liquidity across banks and, in their model, provides greater resilience to systemic risk, while other work shows that concentration around highly connected institutions can amplify or contain contagion depending on the location of shocks, the distribution of exposures, and banks' loss-absorbing capacity \citep{sui2020financial,siebenbrunner2021quantifying}.

More generally, the consequences of interconnectedness depend on the nature of the shocks hitting the system. \cite{georg2013} distinguishes contagion transmitted through interbank links from common shocks and shows that their effects depend differently on network structure. \cite{castiglionesi2019financial} show that financial integration facilitates liquidity reallocation when shocks are uncorrelated, but can increase fragility under correlated shocks by inducing banks to hold fewer liquid resources. Similarly, diversification across a larger number of counterparties need not reduce systemic risk when amplification mechanisms are present \citep{battiston2012default}. A related trade-off arises in \cite{cabrales2017risk}: dense networks improve diversification under light-tailed shocks but become inefficient when shocks are heavy-tailed, making segmentation or assortative matching socially preferable.

More recently, \cite{shu2025endogenous} endogenizes banks’ risk-taking decisions within interbank networks, showing that greater interconnectedness can distort incentives and induce higher ex-ante risk-taking, thereby generating systemic fragility even in the absence of large exogenous shocks. Overall, these theoretical contributions highlight the need to distinguish institutions that amplify shocks from those that absorb them, and to separate structural contagion channels from market-based ones. Our paper examines these distinctions using a model-implied contemporaneous transmission representation and complementary indices of systemic impact and vulnerability. Structural interpretation remains conditional on the recursive non-Gaussian identifying assumptions.

The causal-discovery component builds on a growing body of work that brings explicit causal-discovery methods to economics and finance. The modern toolkit rests on directed-acyclic-graph and structural-causal-model theory \citep{pearl2000causality,pearl2009causality,Peters2017} and on the discovery algorithms surveyed by \cite{glymour2019review}. Within this toolkit, the Linear Non-Gaussian Acyclic Model of \cite{Shimizu06a} established identification of linear acyclic structural models under mutually independent non-Gaussian disturbances, including the orientation of contemporaneous edges, through independent component analysis \citep{comon1994independent,hyvarinen2013independent}; \cite{shimizu2011directlingam} subsequently introduced DirectLiNGAM, a deterministic estimator that avoids the initialization and convergence problems of the ICA-based procedure and which we adopt here. The time-series extension VAR-LiNGAM \citep{hyvarinen2010} separates lagged from contemporaneous effects and has been applied widely in economics, to firm growth and productivity \citep{Moneta_2013,ciarli2023does}, the identification of monetary-policy shocks \citep{moneta2004identification}, commodity price discovery \citep{xu2014}, and linkages across international stock and foreign-exchange markets \citep{jiang2023}. In the specific setting of financial crises and contagion, causal-discovery methods have been used to construct stress-testing scenarios with Suppes--Bayes causal networks \citep{GAO2017,GAO2018} and to map contagion across global currencies \citep{RIGANA2023}, while \cite{kumar2023} provide a broader survey of causal-inference applications in banking, finance, and insurance. Relative to this literature, our contribution is twofold. Substantively, we apply VAR-LiNGAM to a broad cross-section of European banks together with the market index and translate the recovered structure into market-capitalization-weighted indices of systemic impact and vulnerability. Methodologically, we make the \emph{reproducibility} of the discovery output an explicit part of the analysis: we assess the identifying assumptions through independence and recursiveness diagnostics, distinguish the directionally reproducible backbone of the network from links that are sensitive to resampling, and examine how uncertainty in individual directions propagates to the resulting systemic-impact and vulnerability rankings. Following the distinction between persistence and independence support in \citet[Section 3.1]{doannguyenbucciol2026}, our re-discovery bootstrap repeats ordering and coefficient selection in every replication and propagates the resulting uncertainty to SIS/SVS rankings.

To sum up, theoretical models of contagion and empirical approaches to systemic-risk measurement - including market-based measures, econometric connectedness methods, and structural network models - offer complementary insights but also leave important gaps. Market-based indicators provide forward-looking signals of vulnerability and have the advantage of relying on publicly available market data that are timely and relatively less subject to accounting discretion or regulatory manipulation. Connectedness measures capture directional and dynamic spillovers across institutions, but remain based on reduced-form associations. Structural network models, by contrast, provide a more explicit representation of contagion channels, but often require granular information on exposures and typically do not identify contemporaneous transmission directly from market data. Our paper contributes to bridging these perspectives by applying a causal-discovery framework, VAR-LiNGAM, to European banks using publicly available market data. The approach estimates contemporaneous directional linkages under explicit identifying assumptions and uses the resulting total-effect structure, together with bank market capitalization, to construct measures of systemic impact and vulnerability. In this way, we retain the accessibility of market-based approaches while adding a model-dependent representation of contemporaneous paths and their discovery uncertainty.

A further identification issue concerns common conditional volatility. Recovering linearly uncorrelated disturbances does not, by itself, establish the mutual independence required by LiNGAM. Shared stochastic scales can introduce nonlinear dependence even when conventional covariance-based checks look satisfactory \citep{montielolea2022}. In this application, the discovery step is therefore performed after univariate volatility standardization, and its recovered shocks are assessed using a pairwise kernel-dependence diagnostic. The distributional evidence is interpreted conditionally on the model class, with the limitations of pairwise testing, omitted factors and a daily recursive approximation kept distinct from bootstrap reproducibility. This distinction also separates model-implied market-return pathways from balance-sheet contagion channels and from financial price discovery \citep{lo1990,chordia2000,hasbrouck1995}.

\section{Structural causal model and systemic contagion indices}
\label{sec:methods}

\subsection{Directed Acyclic Graph (DAG)}

A DAG is the graphical representation of a structural causal model (SCM). It is a finite graph consisting of nodes and directed edges with no cycles: starting from any node and following the directed edges, one can never return to that node.

In causal reasoning, DAGs are a convenient tool for representing directed relationships between variables. Each node represents a variable, and a link from node \( X \) to node \( Y \) (\( X \rightarrow Y \)) indicates that \( X \) is a direct cause of \( Y \).

Formally, a DAG \( \mathcal{G} = (V, E) \) consists of:
\begin{itemize}
  \item A set of nodes \( V = \{X_1, X_2, \dots, X_n\} \),
  \item A set of directed edges \( E \subseteq V \times V \),
\end{itemize}
with the constraint that there is no directed path \( X_i \rightarrow \cdots \rightarrow X_i \) forming a cycle.

\subsection{Vector AutoRegression (VAR)}

A vector autoregression (VAR) describes joint linear dynamics in multivariate time series. The model is suitable for finding linear dependence between variables over time. Suppose \( \bm{X}_t \in \mathbb{R}^d \) represents a \( d \)-dimensional vector of observed time series variables, the p-order VAR model is expressed as follows:

\begin{equation}
\label{eq:var}
\bm{X}_t = \sum_{l=1}^p \bm{A}_l \bm{X}_{t-l} + \bm{u}_t
\end{equation}

where:
\begin{itemize}
    \item \( A_l \in \mathbb{R}^{d \times d} \) are coefficient matrices that capture the lag-\( l \) influence of past values;
    \item \( \bm{u}_t \in \mathbb{R}^d \) is a white noise vector with covariance matrix $E[\bm{u}_t \bm{u}'_t] = \bm{\Sigma}_u$
\end{itemize}

The VAR model (\ref{eq:var}) explains each variable through its own lagged values and those of the other variables. It is a reduced-form model and does not, on its own, identify contemporaneous causal relationships. The corresponding structural representation --- the Structural VAR model --- can be written as
\begin{equation}
\label{eq:svar}
(\bm{I}-\bm{B}_0)
\left(
\bm{X}_t - \sum_{l=1}^{p}\bm{A}_l\bm{X}_{t-l}
\right)
=
\bm{\varepsilon}_t ,
\end{equation}
where:
\begin{itemize}
    \item \( \bm{A}_l \in \mathbb{R}^{d \times d} \) are the reduced-form VAR coefficient matrices.
    \item \( \bm{B}_0 \in \mathbb{R}^{d \times d} \) is the contemporaneous causal coefficient matrix.
    \item \( \bm{u}_t = \bm{X}_t - \sum_{l=1}^{p}\bm{A}_l\bm{X}_{t-l} \) is the reduced-form VAR residual vector.
    \item \( \bm{\varepsilon}_t = (\bm{I}-\bm{B}_0)\bm{u}_t \) is the vector of %mutually independent, non-Gaussian 
    structural shocks, with
    \( \mathbb{E}[\bm{\varepsilon}_t\bm{\varepsilon}_t'] = \bm{D}_{\varepsilon} \).
\end{itemize}

This structural representation is identical to the VAR-LiNGAM framework (presented below) in the particular case of $\bm{\varepsilon}_t$ mutually independent and non-Gaussian (with at most one exception), and $\bm{B}_0$ lower triangular for at least one specific ordering of the variables.

\subsection{Linear Non-Gaussian Acyclic Model (LiNGAM)}
\label{sec:lingam_est}

The Linear Non-Gaussian Acyclic Model (LiNGAM) is a statistical model for recovering causal structure from data \citep{Shimizu06a}. The LiNGAM model assumes linear contemporaneous relationships, an acyclic contemporaneous structure, and mutually independent exogenous disturbances with at most one Gaussian component. Given the reduced-form residual vector \( \bm{u}_t \in \mathbb{R}^d \), which captures the contemporaneous (same-period) interactions between variables, the LiNGAM model is defined as follows:

\[
\bm{u}_t = \bm{B}_0 \bm{u}_t + \bm{\varepsilon}_t
\Rightarrow
\bm{u}_t = (\bm{I} - \bm{B}_0)^{-1}\bm{\varepsilon}_t .
\]

where:
\begin{itemize}
    \item \( \bm{u}_t \in \mathbb{R}^{d} \) is the reduced-form residual vector obtained from the VAR model.
    \item \( \bm{B}_0 \in \mathbb{R}^{d \times d} \) is the contemporaneous causal coefficient matrix encoding the Directed Acyclic Graph (DAG) structure of causal relationships among variables at the same time point. Since it encodes an acyclic graph, $\bm{B}_0$ is essentially lower triangular (i.e, $\bm{P} \bm{B}_0 \bm{P}^{\prime}$ is lower triangular for some permutation matrix $\bm{P}$).
    \item \( B_{0,ij} \) represents the model-implied contemporaneous effect from variable \( j \) to variable \( i \).
    \item \( \bm{\varepsilon}_t \in \mathbb{R}^{d} \) is a vector of mutually independent, non-Gaussian structural shocks.
\end{itemize}

Provided the model assumptions hold, LiNGAM identifies the contemporaneous structure among the variables. Embedded in the structural VAR, in the VAR-LiNGAM framework, it separates the contemporaneous component from the lagged dynamics.

We estimate $\bm{B}_0$ with the DirectLiNGAM algorithm \citep{shimizu2011directlingam}, which recovers the causal ordering by iteratively selecting, as the next root, the variable that is most independent of the residuals of the others, and then obtains the coefficients by regression along that ordering. Independence is measured with the pairwise likelihood-ratio criterion of \citet{hyvarinen2013pairwise}, and each variable is regressed by ordinary least squares on all variables that precede it in the ordering. DirectLiNGAM uses a deterministic ordering criterion conditional on the input data and, unlike ICA-based estimators, does not depend on random initialization or on the convergence of an iterative unmixing step, which makes it better suited to a system of this dimension. A regression-based fit yields a fully connected lower-triangular $\bm{B}_0$, because every earlier-ordered variable receives a coefficient. All series are standardized to unit variance before the LiNGAM regression, so each entry of $\bm{B}_0$ is a standardized contemporaneous coefficient. The $0.05$ coefficient threshold is used descriptively to display a sparse graph and is also used within the discovery-stability bootstrap to define active directions. Baseline inferential selection, however, is performed over the complete set of order-allowed coefficients rather than over coefficients pre-screened by magnitude. The bootstrap Wald procedure of Section~\ref{sec:wald} therefore tests all $d(d-1)/2$ order-allowed coefficients, applies the multiple-testing correction to that full family, and imposes the materiality threshold $|B_{0,ij}|\ge0.05$ only after multiple-testing correction.

\subsection{Vector
Autoregressive Linear Non-Gaussian Acyclic Model (VAR-LiNGAM)}

 We consider a multivariate time series $\bm{X}_t \in \mathbb{R}^d$, whose components are $x_{1t}, x_{2t}, \dots, x_{dt}$, where \( d \) is the number of observed variables (in this paper, $\bm{X}_t$ collects the log-returns of the banks and the market proxy). 

Here is the formula of the VAR-LiNGAM model:

\[
\bm{X}_t
=
\underbrace{\sum_{l=1}^p \bm{A}_l \bm{X}_{t-l}}_{\text{VAR}}
+
\bm{u}_t,
\quad
\bm{u}_t
=
\bm{B}_0 \bm{u}_t
+
\bm{\varepsilon}_t
\Rightarrow
\bm{u}_t
=
\underbrace{(\bm{I} - \bm{B}_0)^{-1}\bm{\varepsilon}_t}_{\text{LiNGAM}} .
\]

where:
\begin{itemize}
    \item \( \bm{A}_l \in \mathbb{R}^{d \times d} \) are the VAR coefficient matrices at lag \( l \).
    \item \( \bm{u}_t \in \mathbb{R}^{d} \) is the reduced-form residual vector capturing contemporaneous interactions not explained by lagged variables.
    \item \( \bm{B}_0 \in \mathbb{R}^{d \times d} \) is the contemporaneous causal coefficient matrix representing a Directed Acyclic Graph (DAG) of instantaneous causal effects.
    \item \( B_{0,ij} \) represents the model-implied contemporaneous effect from variable \( j \) to variable \( i \).
    \item \( \bm{\varepsilon}_t \in \mathbb{R}^{d} \) is a vector of mutually independent, non-Gaussian structural shocks.
\end{itemize}

The VAR-LiNGAM model combines the reduced-form VAR with LiNGAM, so that the two components capture, respectively, the lagged and the contemporaneous structure of the multivariate series. Relative to a traditional VAR, this allows the contemporaneous transmission structure to be identified rather than left unrestricted. Identification rests on two assumptions. First, the structural shocks \( \bm{\varepsilon}_t \) must be non-Gaussian and mutually independent. In the original ICA-based VAR-LiNGAM framework \citep{comon1994independent,hyvarinen2013independent}, this non-Gaussianity is what allows the contemporaneous ordering to be recovered through Independent Component Analysis; in this paper we implement the LiNGAM step with DirectLiNGAM \citep{shimizu2011directlingam}, which instead recovers the ordering through sequential independence comparisons and regression residualization (Section~\ref{sec:lingam_est}). Second, that structure must be acyclic, with no contemporaneous feedback loops. Under these assumptions the matrix \( \bm{B}_0 \) is uniquely identified and estimable. Representing \( \bm{B}_0 \) as a DAG, with each variable a node and each non-zero entry a directed edge, provides a compact description of the estimated contemporaneous relationships, the validity of which depends on the maintained assumptions holding in the data.

The general representation nests a no-lag specification. With an intercept or mean vector $\bm\mu$, setting $p=0$ gives
\begin{equation}
\bm X_t-\bm\mu=\bm u_t,\qquad
\bm u_t=\bm B_0\bm u_t+\bm\varepsilon_t.
\label{eq:var0_boundary}
\end{equation}
Thus, VAR$(0)$ is the contemporaneous boundary case of VAR-LiNGAM and is estimated by applying DirectLiNGAM to mean-centered, volatility-standardized returns. This is the baseline selected in the empirical application. The VAR$(1)$ implementation applies DirectLiNGAM to standardized one-lag reduced-form innovations and is reported as a robustness analysis. The distinction matters for interpretation: the baseline maps same-day structure without estimating a high-dimensional lag matrix, whereas the robustness specification first removes linear one-day predictive dependence.

\subsubsection{Total contemporaneous transmission matrix}
The total contemporaneous transmission matrix is:
\[
    \bm{C} = (\bm{I} - \bm{B}_0)^{-1}.
\]
This formulation captures both direct model-implied transmission and indirect transmission propagated through intermediate nodes. In other words, \( C_{ij} \) measures the total contemporaneous model-implied transmission from a structural shock to bank \( j \) to bank \( i \), aggregating both immediate and propagated channels.

The empirical implementation distinguishes the unrestricted coefficient matrix used to recover shocks from the statistically selected network used to calculate total effects and SIS/SVS. The independence condition is assessed for the unrestricted representation: it would be inappropriate to treat pruning as a means of repairing failed identification. Because the daily panel exhibits conditional heteroskedasticity, the baseline applies the discovery algorithm to filtered and subsequently standardized returns, as detailed next. This preprocessing choice changes the scale of the structural sensitivity estimand and is not assumed to be an identification theorem.

\subsubsection{Conditional volatility and the scale of the estimand}
\label{sec:volatility}
Shared volatility can make shocks uncorrelated without making them independent \citep{montielolea2022}. In the shared-scale class studied by \citet[Lemma A.2]{doannguyenbucciol2026}, dependence survives every non-singular linear recovery. Whether a volatility standardization applied before recovery removes the detectable part is an empirical question, answered in Section~\ref{sec:independence_results}. We therefore fit a univariate GARCH(1,1) with Student-$t$ innovations to each centered return and divide by its fitted conditional standard deviation. These filtered series are subsequently centered and standardized to unit variance; write their vector as $Z_t$. The baseline discovery step applies to $Z_t$. The VAR(1) sensitivity is fitted to $Z_t$ and applies discovery to its standardized innovations. The unfiltered return model is reported separately.

Filtering changes the scale on which the model is estimated. It is not an identification theorem. If a raw-return model has a constant coefficient matrix $B^{r}$ and $Z_t=D_t^{-1}r_t$ for a time-varying diagonal scale $D_t$, then
\begin{equation}
 Z_t=(D_t^{-1}B^{r}D_t)Z_t+D_t^{-1}\varepsilon_t.
 \label{eq:scale}
\end{equation}
The transformed coefficient matrix is generally time-varying. Positive diagonal scaling preserves zeros and signs at a given date, but does not ensure that a fixed matrix estimated after univariate filtering equals a fixed raw-return structural matrix. The application therefore estimates a recursive approximation on the volatility-adjusted scale, and reassesses residual dependence on that scale. The simulation exercise evaluates selected finite-sample designs; it does not remove this qualification.

The baseline target is a contemporaneous sensitivity on the volatility-adjusted return scale. The unfiltered specification is retained to show how the estimated roles depend on this input transformation. A lower number of independence rejections is not, by itself, a reason to interpret the filtered network as the true raw-return structure. Both specifications are therefore reported with their own recovered-shock diagnostics.

GARCH parameters are estimated by constrained maximum likelihood with a variance-standardized Student-$t$ density. The mean is fitted before the variance recursion, and the filtered observations are scaled using sample moments. In the temporal holdout exercise, all parameter estimates and scaling moments use the training window only. Observations are indexed by retained rows after the calendar screen; conditional-volatility dynamics are consequently an observation-time approximation rather than a model of every intervening exchange session.

\subsection{Independence test - Hilbert Schmidt Independence Criterion (HSIC)}
\label{sec:hsic}
Sections~\ref{sec:hsic}--\ref{sec:tail_method} contain the post-estimation diagnostics. They are organised according to the post-recovery protocol of \citet{doannguyenbucciol2026}, in which empirical support for the identifying independence assumption (Section~\ref{sec:hsic}) and the persistence of recovered relations under re-discovery (Section~\ref{sec:ordering_stability}) are distinct, non-nested layers of evidence, with tail co-exceedance as a complementary diagnostic (Section~\ref{sec:tail_method}); where the present implementation departs from that protocol, the departure is stated at the point of use.

An important next step in VAR-LiNGAM is to test the mutual independence of the structural shocks \( \bm{\varepsilon}_t \), which is the condition that makes the recovered contemporaneous structure identifiable. We treat this as a diagnostic to be tested and reported rather than assumed: we employ the Hilbert--Schmidt Independence Criterion (HSIC) \citep{Gretton_2005_HSIC,Gretton_2007_HSIC_Test}, a nonparametric kernel-based measure of statistical dependence, applied pairwise to the recovered shocks. In the protocol of \citet[Section 3.2 and Table 1]{doannguyenbucciol2026}, independence support for the recovered shocks is the core evidentiary basis for structural credibility: non-Gaussianity opens the distribution-based route and the recursive form restricts it, but neither certifies independence, which is assessed after recovery. Unlike correlation-based measures, HSIC is capable of detecting general nonlinear and higher-order dependence structures between random variables, which matters for financial data whose shocks may be linearly weakly correlated yet dependent through volatility. Section~\ref{sec:independence_results} reports the rejections under the filtered and unfiltered specifications; the diagnostic does not turn a conditional identification assumption into an empirically established fact.

 In theory, two random variables $X$ and $Y$ are independent if and only if
\begin{equation}
P_{XY} = P_X P_Y.
\end{equation}

Directly verifying this condition is infeasible, as it involves infinitely many transformations. HSIC addresses this problem by embedding the variables into reproducing kernel Hilbert spaces (RKHS), where dependence can be assessed through covariance operators.

\subsubsection{Kernel embedding.} Let $k(\cdot,\cdot)$ and $l(\cdot,\cdot)$ denote positive definite kernels associated with the feature maps
\begin{equation}
\phi(X) \in \mathcal{H}, \quad \psi(Y) \in \mathcal{G}.
\end{equation}
The cross-covariance operator between the embedded variables is defined as
\begin{equation}
C_{XY} = E[(\phi(X) - \mu_X) \otimes (\psi(Y) - \mu_Y)], 
\end{equation}
where $\mu_X = E[\phi(X)]$ and $\mu_Y = E[\psi(Y)]$.
HSIC is then defined as the squared Hilbert--Schmidt norm of this operator:
\begin{equation}
HSIC(X,Y) = \| C_{XY} \|_{HS}^2. 
\end{equation}
If characteristic kernels are used (for example, Gaussian radial basis function kernels), then
\begin{equation}
HSIC(X,Y) = 0 \Longleftrightarrow X \perp Y.
\end{equation}
This property ensures that HSIC can detect general forms of dependence, not only linear correlation.
\textbf{Sample estimator.} Given observations $\{(x_i,y_i)\}_{i=1}^n$, define the kernel matrices
\begin{equation}
K_{ij} = k(x_i,x_j), \quad L_{ij} = l(y_i,y_j),
\end{equation}
and the centering matrix
\begin{equation}
H = I - \frac{1}{n} \mathbf{1}\mathbf{1}'.
\end{equation}
The empirical HSIC estimator is
\begin{equation}
\widehat{\mathrm{HSIC}} = \frac{1}{(n-1)^2} \, \mathrm{tr}(KHLH).
\end{equation}
This statistic measures the degree of correlation between similar samples in $X$ and similar samples in $Y$. If the two variables are independent, the central kernel matrix will be approximately uncorrelated, and the trace term will approach zero.
Within the LiNGAM framework, the structural interpretation of the contemporaneous causal matrix $B_0$ relies on the independence of the recovered structural shocks. Since covariance-based diagnostics are insufficient in non-Gaussian settings, HSIC provides a useful post-estimation diagnostic. By examining whether the recovered shocks exhibit statistically significant pairwise dependence, we obtain supportive evidence on the plausibility of the independence assumption in the data. This diagnostic should not be interpreted as a proof of full joint mutual independence, since pairwise independence is weaker than joint independence. For the empirical implementation we use the exact Gaussian-kernel statistic, with the median pairwise-distance bandwidth on each standardized margin. The disturbances are recovered from the unpruned fitted representation, $\widehat\varepsilon_t=(I-\widehat B_0)Z_t$, because the identification assumption pertains to that representation. A Gaussian kernel is characteristic, so population HSIC vanishes under independence, while sample non-rejection is not proof of independence. Throughout, ``independence support'' means compatibility with the independence condition under this diagnostic at this level, not a verified assumption.

To account for the time-series structure, we use all admissible circular shifts with a minimum circular distance of 20 observations. This choice follows the principle that recovered-shock independence diagnostics require a serial-aware reference, as emphasized by \citet[Section 4.3]{doannguyenbucciol2026}. The circular-shift construction and its calibration are specific to the present application and are not claimed to yield exact inference under unrestricted temporal dependence. The shift reference distribution retains the within-series order while breaking contemporaneous alignment. Let $T_s$ denote the HSIC statistic at admissible shift $s$ and $N$ the number of such shifts; the reference p-value is
\begin{equation}
 \widehat p=\frac{1+\sum_{s=1}^N\mathbf{1}\{T_s\ge T_{\mathrm{obs}}\}}{N+1}.
 \label{eq:shiftp}
\end{equation}
The exact-kernel statistics are computed across admissible circular shifts using the identity in Appendix~\ref{app:implementation}. Benjamini--Hochberg (BH) adjustment at 5\% is the designated pairwise classifier; Benjamini--Yekutieli (BY) adjustment and normalized HSIC magnitudes are reported as sensitivity and severity summaries. Following \citet[Definition 3.3 and Section 4.3]{doannguyenbucciol2026}, non-rejection under the designated classifier is termed operational pairwise independence support. It is neither proof of joint mutual independence nor an unconditional finite-sample error-control guarantee: the interpretation remains conditional on the serial-aware reference and its calibration. BY non-rejection does not override a BH rejection. No links are pruned to remove pairs that reject.

\subsection{Significance test -- Bootstrap Wald test}
\label{sec:wald}
We use a moving-block coefficient bootstrap conditional on the full-sample order. For each order-allowed coefficient, the statistic is
\begin{equation}
 W_{ij}=\widehat B_{0,ij}^{\,2}/\widehat{\operatorname{SE}}_{ij}^{\,2}.
 \label{eq:wald}
\end{equation}
The point estimate and the bootstrap coefficients come from the same predecessor regressions on the full-sample standardized input; the resamples are centered but not re-standardized. We draw 1,000 resamples of blocks of ten consecutive retained observations and use their coefficient standard deviation as the standard error. Wald $p$-values use a one-degree-of-freedom chi-squared reference.

The recovered order allows $d(d-1)/2$ coefficients. BH adjustment at $q=0.05$ is applied to this full family, before the materiality restriction $|\widehat B_{0,ij}|\ge0.05$. Let $\widehat B_0^s$ denote the matrix retained by both rules. This matrix, rather than the unrestricted predecessor matrix, is used for systemic scores. BY selection is also summarized. The procedure is an order-conditional significance assessment, not a general post-selection guarantee for directions selected by the same data. Blocks preserve local dependence in retained-observation time; they do not reconstruct missing trading sessions.

\subsection{Causal Ordering Stability}
\label{sec:ordering_stability}
The Wald test fixes the ordering and therefore does not measure discovery uncertainty. An early position gives a node more potential downstream destinations, so order instability can affect path-based scores even when individual coefficients are precisely estimated conditional on that order.

We consequently conduct 500 moving-block re-discovery replications. Each resample re-estimates the DirectLiNGAM order, repeats predecessor regressions, and applies a new conditional Wald--BH--materiality selection with 200 inner block resamples; the volatility filter is held fixed at its full-sample fit across replications. Systemic scores are recomputed from the resulting sparse graph. For $i\ne j$, the directional frequency is
\begin{equation}
 \widehat P(i\rightarrow j)=\frac{1}{500}\sum_b
 \mathbf{1}\{\widehat B_{0,ji}^{s,(b)}\ne0\}.
 \label{eq:frequency}
\end{equation}
The strict backbone retains baseline links selected in the same source--target direction in at least 80\% of the 500 re-discovery replications. This direction-frequency criterion is motivated by, but distinct from, the persistence definition in \citet[Definition 3.2]{doannguyenbucciol2026}. Each replication re-estimates the ordering and applies conditional Wald--BH and coefficient-materiality selection. No additional cross-replication coefficient-sign-consistency criterion is imposed. The backbone therefore measures reproducibility of selected orientations under the present discovery procedure rather than persistence under all conditions of Definition 3.2. Among unordered pairs whose summed directional frequencies reach 10\%, we identify pairs with both orientations appearing in at least 20\% of replications as ambiguous. These are descriptive thresholds, not posterior probabilities.

We report distributions of topological position and SIS/SVS ranks, their 5th--95th percentile ranges, and top-five inclusion frequencies. Rank ranges describe uncertainty across fitted representations; they are not simultaneous confidence sets. The re-discovery procedure varies ordering and coefficient selection conditional on the initially estimated volatility filter, the full-sample standardization and the retained sample. Each replication resamples rows of the standardized input; the ordering step standardizes internally, whereas coefficients, materiality and scores are computed on the resampled rows without re-standardization. It does not incorporate uncertainty in GARCH parameter estimation, observation selection, hidden common causes, or the maintained model class. The 500 outer replications represent a computationally bounded assessment, not a theoretically mandated bootstrap size. In particular, estimated frequencies near the 80\% backbone cutoff have appreciable Monte Carlo uncertainty. Reported inclusion frequencies should be read at modest precision. Related bootstrap approaches are discussed by \citet{komatsu2010,Moneta_2013}.

Two further influence analyses remove each calendar quarter in turn and each of the fifteen observations with the largest absolute standardized return. For these exercises the filter and discovery step are refitted, and the conditional selection uses 500 bootstrap replications. They assess whether individual observations or episodes dominate the aggregate rankings.

\subsection{Tail Co-Exceedance Diagnostics}
\label{sec:tail_method}
Co-exceedances describe simultaneous extreme market movements \citep{bae2003coexceedance,baur2005coexceedances,chiu2015industry}. We use them to evaluate the tail dependence associated with an estimated graph. Let
\begin{equation}
 D_{i,t}=\mathbf{1}\{Z_{i,t}\le q_i(0.05)\},\qquad
 S_{i,t}=\mathbf{1}\{\widehat\varepsilon_{i,t}\le q_i^\varepsilon(0.05)\}.
 \label{eq:tailindicators}
\end{equation}
The symmetric lift and the shock-conditioned lift are
\begin{equation}
 \operatorname{CoLift}_{ij}=\frac{P(D_i=1,D_j=1)}{P(D_i=1)P(D_j=1)},\qquad
 \operatorname{DirLift}_{i\rightarrow j}=\frac{P(D_j=1\mid S_i=1)}{P(D_j=1)}.
 \label{eq:tail}
\end{equation}
Symmetric lift compares linked unordered pairs with genuine non-links. Directional lift compares the estimated direction with the reverse direction for the same pair, and with genuine non-links respecting the recovered order. An edge and its reverse are never counted as a genuine non-link. The system-level diagnostic includes the market node; we additionally report bank-to-bank edges separately.

A node-label permutation keeps the graph topology fixed and relabels its endpoints on the already-computed lift matrices. With 1,000 permutations, the smallest attainable reference $p$-value is $1/1001$. This answers a narrow question about alignment between graph labels and a fixed tail-dependence representation. Because shocks themselves depend on the recovered order, the permutation does not establish that the selected direction is superior to directions obtained from other recursive models.

\subsubsection{Temporal holdout design}
\label{sec:holdout_method}
We estimate the volatility parameters, centering and scaling moments, ordering, coefficient matrix, sparse graph, and 5\% tail thresholds using observations through December 2021. The variance recursion then continues through the later observations with those parameters held fixed. The test-period disturbance is
\begin{equation}
 \widetilde\varepsilon_t^{\mathrm{test}}
 =(I-\widehat B_0^{\mathrm{train}})Z_t^{\mathrm{test}}.
 \label{eq:holdout}
\end{equation}
We call this a \emph{training-model recovered disturbance}; it is a structural shock for the later period only if the relevant structural mapping remains adequate. No test-period fitting is used for the main holdout statistic. The evaluation necessarily uses realized test observations and is conditional on the selected universe and processing rules. It is not a real-time trading backtest. The split is a retrospective diagnostic with analysis choices informed by the broader sample, not a preregistered external validation.

Both events in equation~\eqref{eq:tail} occur at time $t$. The exercise therefore examines the transport of a same-period tail-dependence pattern to later data, rather than forecasting $Z_{t+1}$ or observing an intervention. We obtain a percentile interval for the forward-minus-reverse gap by resampling test dates in blocks of ten with the fitted transformation and thresholds fixed. We also exclude sessions on which ten or twenty bank indicators co-exceed as stress-day sensitivities.

\subsubsection{A comparator that re-estimates the shock transformation}
\label{sec:coherent}
A positive holdout gap could arise from the recursive residualization itself. To examine this possibility, we generate 200 alternative orders, keeping the market node first and randomly permuting the bank nodes. For each order, we re-estimate all predecessor regressions on the training window, repeat conditional Wald--BH--materiality selection with 1,000 inner replications, construct that order's own training shock thresholds, and apply its own transformation to the same test observations. Volatility filtering and return thresholds remain common across specifications.

This comparison changes both the graph and the recovered shocks. It is consequently more demanding than relabeling an unchanged lift matrix, although differing selected graph densities mean it is a diagnostic rather than an exact randomization test. Appendix~\ref{app:counterexample} gives a simple common-cause example showing why either recursive direction can yield a positive held-out gap. Such an example does not challenge LiNGAM's identification theorem under its assumptions; it limits what the gap alone can establish about those assumptions or about direction.

\subsection{Systemic risk indices}
Given a vector of market capitalizations $m = [m_1, m_2, \dots, m_d]^\top$, combined with the total-effect matrix $C$, we propose two systemic risk indices. These indicators should be interpreted as market-capitalization-weighted return sensitivity measures based on the estimated contemporaneous structure. Integrating market capitalization helps combine the estimated transmission strength with the economic size of the affected or originating banks.

The baseline SIS and SVS are computed from the selected bank-only total-effect matrix $\bm{C}^{B}=(\bm{I}-\widehat{\bm{B}}^{s}_{0,BB})^{-1}$. In the empirical analysis, $\bm{B}_0$ denotes the full-sample contemporaneous matrix after full-family Benjamini--Hochberg correction and the post-inference $0.05$ materiality threshold, so that the indices are functions of the coefficient-selected baseline network rather than of the unpruned lower-triangular regression matrix. We construct the bank-only total-effect matrix from the selected bank-to-bank coefficient block, excluding the market proxy from the reported bank-level scores. In the baseline recovered ordering, the market proxy is first and has no incoming bank links. Consequently, the bank-to-bank block of the full inverse coincides with the inverse of the bank-only block. This equality need not hold for alternative recursive orderings in which the market node lies on paths connecting banks. Neither this convention nor the presence of the market node eliminates omitted common causes. The bootstrap directional frequencies introduced in Section~\ref{sec:ordering_stability} assess the reproducibility of selected orientations rather than replacing their coefficients. The SIS/SVS definitions below consequently use the full-sample selected bank-only matrix; their re-discovery distributions are reported separately.

\subsubsection{Systemic Impact Score (SIS)}

The Systemic Impact Score (SIS) measures the market-capitalization-weighted sensitivity of the banking system to a structural shock originating from a bank. It accounts for all paths of influence in the total-effect matrix and weights the affected nodes by their market size:
Let \( \mathcal{B} \) denote the set of bank nodes. The market proxy node is excluded from the market-capitalization-weighted SIS and SVS calculations because it does not have a bank-level market capitalization.

\[
\text{SIS}_j = \sum_{\substack{i \in \mathcal{B} \\ i \neq j}} |C^B_{ij}| \cdot m_i ,
\qquad j \in \mathcal{B}.
\]

The higher the SIS index, the greater the estimated influence of the bank on the rest of the system. Since $C^B_{ij}$ is an effect on filtered returns, SIS should be interpreted as a relative systemic impact measure. A direct interpretation as a monetary loss requires specifying the size, sign, and horizon of the shock.

Taking the absolute value $|C^B_{ij}|$ treats amplifying and dampening transmission symmetrically, which is appropriate when the quantity of interest is the magnitude of transmission. The total effects need not all be positive, however: a negative $C^B_{ij}$ means that a positive shock to bank $j$ is associated, through the estimated network, with a downward effect on bank $i$. A negative total effect represents a model-implied offsetting contemporaneous sensitivity. It does not, by itself, identify an economic hedging or substitution mechanism. To keep this distinction visible we decompose SIS into an amplifying and a dampening part,
\begin{equation}
\text{SIS}^{+}_j = \!\!\sum_{\substack{i \in \mathcal{B},\, i\neq j \\ C^B_{ij}>0}}\!\! C^B_{ij}\, m_i,
\qquad
\text{SIS}^{-}_j = \!\!\sum_{\substack{i \in \mathcal{B},\, i\neq j \\ C^B_{ij}<0}}\!\! |C^B_{ij}|\, m_i,
\qquad
\text{SIS}_j = \text{SIS}^{+}_j + \text{SIS}^{-}_j,
\label{eq:signed_sis}
\end{equation}
and report the share $\text{SIS}^{-}_j/\text{SIS}_j$ carried by dampening effects as a robustness check on the magnitude-based ranking.

\subsubsection{Systemic Vulnerability Score (SVS)}

The Systemic Vulnerability Score measures how strongly bank $i$ is exposed to shocks from other banks in the estimated network. It captures the total influence it receives from others, weighted by the market size of the originator banks:
\[
\text{SVS}_i = \sum_{\substack{j \in \mathcal{B} \\ j \neq i}} |C^B_{ij}| \cdot m_j ,
\qquad i \in \mathcal{B}.
\]

A high SVS suggests that bank $i$ is highly exposed to shocks from the rest of the network. As with SIS, SVS is a market-capitalization-weighted return sensitivity measure rather than a realized loss estimate unless a specific shock scenario is defined.

When taking into account the market capitalization of banks, we define the scale-adjusted SVS (nSVS) as its vulnerability relative to its own market capitalization:

\[
\text{nSVS}_i = \frac{\text{SVS}_i}{m_i},
\qquad i \in \mathcal{B}.
\]

This scaling distinguishes the model-implied incoming sensitivity relative to a bank's own size from the absolute level of incoming sensitivity; high nSVS need not imply a large aggregate system-wide loss.

\subsubsection{Summary of systemic risk indices}
\begin{itemize}

\item \textbf{SIS}: a relative measure of the market-capitalization-weighted sensitivity of the banking system to shocks originating from a given bank.
\item \textbf{SVS}: a relative measure of a bank's exposure to shocks originating from the rest of the estimated network.
\item \textbf{nSVS}: a scale-adjusted vulnerability measure obtained by normalizing SVS by the bank's own market capitalization.
\end{itemize}
Taken together, these indices give a more complete picture of each bank's position in the estimated network.

\subsubsection{Size-controlled benchmark correlations}
Because SIS and SVS explicitly incorporate market capitalization, part of their correlation with external benchmarks may reflect a common association with bank size. We therefore report, only as a benchmark-comparison diagnostic, first-order partial correlations that control for market capitalization. Let $X$ denote a model-implied transmission score (SIS or SVS), $Y$ an external benchmark (Diebold--Yilmaz connectedness or MES), and $Z$ bank market capitalization. The first-order partial correlation is
\begin{equation}
\rho_{XY\cdot Z}
=
\frac{\rho_{XY}-\rho_{XZ}\rho_{YZ}}
{\sqrt{\left(1-\rho_{XZ}^{2}\right)\left(1-\rho_{YZ}^{2}\right)}}.
\label{eq:partial_corr_cap}
\end{equation}
For the rank-based version, each pairwise $\rho$ in Equation~\eqref{eq:partial_corr_cap} is a Spearman correlation; for the linear version, each is a Pearson correlation on the raw variables. Equivalently, the statistic correlates the residual components of $X$ and $Y$ after their separate first-order associations with $Z$ have been removed. This calculation is used only to assess whether benchmark agreement survives a simple size control. It does not alter the definitions or rankings of SIS, SVS or nSVS, does not establish causality, and does not control for other potential common determinants.

The empirical benchmark additionally controls for the loading on the first principal component of the standardized bank returns. In that case the reported partial rank correlation is computed from the correlation between residuals of the separately ranked quantities after regression on an intercept and the ranked size and common-loading controls. Neither operation establishes incremental predictive usefulness of the systemic scores.

\subsubsection{Relationship to Diebold--Yilmaz connectedness and MES}
\label{sec:bench_method}
We estimate a VAR(1) for standardized bank returns and use a generalized forecast-error variance decomposition at horizon $H=10$ \citep{pesaran1998generalized,diebold2014network}. Let $\Phi_h$ denote its moving-average coefficients and $\Sigma$ its innovation covariance. The unnormalized variance share is
\begin{equation}
 \theta_{ij}(H)=\frac{\sigma_{jj}^{-1}\sum_{h=0}^{H-1}(e_i'\Phi_h\Sigma e_j)^2}
 {\sum_{h=0}^{H-1}e_i'\Phi_h\Sigma\Phi_h'e_i}.
\end{equation}
We normalize each row to sum to one. Off-diagonal column sums give ``to others'' connectedness, and off-diagonal row sums give ``from others'' connectedness. Individual ``to'' totals need not be bounded by 100\%; total connectedness divides the sum of off-diagonal shares by the number of banks. The benchmark here is return-based and uses the same closing-price panel as the structural model.

MES is the negative mean raw bank log return on the worst 5\% of raw market-proxy return observations, in a positive-loss convention \citep{acharya2017}. It measures exposure to market distress, not bilateral transmission pathways. SIS/SVS, generalized connectedness, and MES have different conditioning and scaling conventions; their comparison is descriptive rather than a competition among interchangeable estimators.

\section{Data}
We work with a predefined universe of listed institutions and a broad equity-market proxy. Our observation and volatility procedures address the comparability of recorded return intervals and the scale of the input to contemporaneous discovery; they do not select institutions according to their estimated systemic-impact or vulnerability ranks.

\label{sec:data}
\subsection{Sample and market variables}
The panel contains 51 listed banks from sixteen European countries and EXSA.DE, a STOXX Europe 600 tracker, as a market-wide reference node. The bank list and market capitalizations are reported in Appendix Table~\ref{tab:banks}. Italy contributes fourteen institutions; Spain and Poland five each; Germany and Greece four each; Denmark and France three each; Austria, Ireland, the Netherlands, and Sweden two each; and Belgium, Finland, Portugal, the Czech Republic, and Hungary one each.

The data contain 2,205 aligned closing-price observations from 20 November 2015 to 27 June 2025. Prices are daily adjusted closes from Yahoo Finance. We take logarithmic first differences, obtaining 2,204 return intervals. Market capitalizations are taken from Yahoo Finance as of 1 July 2025 and expressed in EUR billion. Their sum is EUR 1,191.449 billion. No denominator for European banking-system asset or market-capitalization coverage is available, so we do not infer one.

This is a selected listed-bank panel, not a census of European banks and not a reconstructed historical entry-and-exit universe. Failed or absorbed institutions absent from the specified universe are not added retrospectively. Conclusions therefore concern the included listed equities. In particular, the exercise cannot directly evaluate which institutions would have transmitted losses before their failure or disappearance from the panel.

\subsection{Return intervals and the calendar screen}
\label{sec:cleaning}
Synchronizing price levels by intersecting trading dates can create multi-session return intervals. If one exchange is closed, the next date retained in the joint price panel may span more than one eligible business day for another exchange. The resulting return does not necessarily measure one common trading session. Such timing issues are closely related to asynchronous-return problems discussed by \citet{burns1998,martens2001}.

Our baseline begins in January 2016 and applies an observable calendar screen to the aligned price panel. An interval is excluded when it contains an intervening eligible weekday, with weekends, 1 January, Good Friday, Easter Monday, 1 May, and 25--26 December omitted from the eligible-day calendar. Table~\ref{tab:sample} reports the observation accounting. After excluding 24 pre-2016 intervals and 190 post-2015 intervals failing this rule, the baseline contains \Nobs{} observations, from 5 January 2016 to 27 June 2025.

\begin{table}[!htbp]
\centering
\caption{Observation accounting and the baseline sample.}\label{tab:sample}
\small
\begin{tabular}{p{11.4cm}r}
\toprule
Step & Observations \\
\midrule
Aligned price rows & 2,205 \\
Log-return intervals & 2,204 \\
Intervals dated before January 2016 & 24 \\
Post-2015 intervals failing the calendar screen & 190 \\
Calendar-screened baseline (no return-magnitude cutoff) & 1,990 \\
\bottomrule
\end{tabular}
\notetext{The calendar rule counts intervening eligible weekdays after removing the common holidays stated in the text. It is a screen on the aligned panel, not a verified reconstruction of each exchange calendar. Return-magnitude omissions are assessed separately as sensitivities, not imposed in the baseline.}
\end{table}

This rule improves comparability of recorded intervals but does not certify that each remaining return is a genuine single-session return for every stock. The price panel does not contain exchange-specific trading calendars, individual-stock suspension flags, adjustment factors, or the underlying unsynchronized price histories. A stale or suspended stock can therefore pass the common-calendar screen. We therefore report issuer-exclusion and alternative-calendar specifications as sensitivities rather than treating the screen as a complete reconstruction of individual trading sessions. After an interval is omitted, the next retained value remains its original adjacent-price return; we do not recompute returns across the newly created gap.

The January 2016 start is an analysis-window choice rather than a claim that all earlier returns are invalid. A sensitivity extends the calendar-screened sample back to the available 2015 observations. Another retains all post-2015 aligned intervals without the calendar screen. These comparisons expose the consequences of the choices instead of treating a particular date boundary as self-validating.

\subsection{Volatility standardization and empirical diagnostics}
\label{sec:large_returns}
The baseline retains observed large market movements rather than excluding dates solely by an absolute-return threshold. We assess sensitivity to return-magnitude screens and to the exclusion of the two institutions with the largest discontinuities in Section~\ref{sec:var1_robustness}. These screens do not replace verification of stock-specific suspensions or corporate-action adjustments, which is not possible from the synchronized closing-price panel alone.

Following Section~\ref{sec:volatility}, the baseline fits univariate GARCH(1,1) models with Student-$t$ innovations before contemporaneous discovery. The model is estimated separately for each of the 52 centered return series. Fitted conditional standard deviations are used to filter each series, after which the filtered observations are centered and scaled to unit variance. All 52 univariate GARCH(1,1)-$t$ models converge successfully under the final estimation settings. Median estimates are $\alpha=0.082$, $\beta=0.895$, and $\nu=4.72$. Some persistence estimates reach the unit boundary: the variance recursion remains usable as a filter, but this does not imply a finite unconditional variance.

Filtering reduces median excess kurtosis from 7.85 to 3.76, while leaving all input series non-Gaussian according to the reported distributional diagnostics. This improvement is cross-sectional, not uniform: extreme standardized observations and potentially irregular issuer histories remain. In particular, the maximum filtered excess kurtosis is about 1,859, highlighting the limits of univariate volatility filtering. Filtering is therefore treated as an empirical preprocessing step rather than a guarantee of independent shocks or a proof that the raw-return structural coefficient matrix is preserved. Section~\ref{sec:independence_results} evaluates the remaining recovered-shock dependence, Section~\ref{sec:var1_robustness} reports sample and issuer sensitivities, and Appendix~\ref{app:mc} describes a controlled simulation comparison.

\section{Empirical Results}
\label{sec:results}
\subsection{Assumption checks and empirical lag specification}
The augmented Dickey--Fuller tests reject a unit root for all standardized series. Shapiro--Wilk and Jarque--Bera tests reject marginal Gaussianity for all 52 filtered inputs; Jarque--Bera also rejects for all unpruned recovered disturbances. These are distributional diagnostics, not proofs of temporal stability or independent structural errors. The dependence assessment is reported separately in Section~\ref{sec:independence_results}. These diagnostics address different questions: stationarity and non-Gaussianity concern the suitability of the observed series for the proposed estimator; recovered-shock dependence probes a central identifying assumption; and direction-frequency re-discovery assesses the reproducibility of selected orientations. None alone verifies the full structural model or causal sufficiency, and none substitutes for another.

\begin{table}[!htbp]
\centering
\caption{Lag-order selection on a common dependent-observation sample.}\label{tab:lagselect}
\small
\begin{tabular}{lrrrr}
\toprule
Lag & AIC & BIC & HQ & FPE ($10^{-16}$) \\
\midrule
0 & -36.657 & -36.510 & -36.603 & 1.203 \\
1 & -35.785 & -28.019 & -32.932 & 2.879 \\
2 & -34.690 & -19.305 & -29.039 & 8.637 \\
3 & -33.579 & -10.574 & -25.128 & 26.578 \\
4 & -32.438 & -1.814 & -21.190 & 85.083 \\
5 & -31.325 & 6.918 & -17.278 & 269.262 \\
\bottomrule
\end{tabular}
\notetext{All criteria are computed for the 52-variable standardized panel with an intercept and maximum lag five. The common dependent-observation sample has 1,985 rows. Lags refer to retained-observation time.}
\end{table}

Table~\ref{tab:lagselect} compares lag orders zero through five on the same dependent-observation sample. AIC, BIC, HQ, and FPE all select $p=0$. We therefore retain the contemporaneous specification as the baseline. This choice reflects the trade-off between fit and the number of lag coefficients; it does not establish that all lagged dependence is absent. Ljung--Box tests reject at lag five for 30.8\% of filtered series in levels and 23.1\% in squares, compared with 98.1\% for squared unfiltered returns. The VAR(1) sensitivity in Section~\ref{sec:var1_robustness} consequently remains informative.

\subsection{Estimated contemporaneous network among European banks}
\label{sec:network_results}
DirectLiNGAM recovers a recursive ordering and 1,326 order-allowed predecessor coefficients. A descriptive threshold of 0.05 leaves \Nmaterial{} links. Figure~\ref{fig:full_network} preserves the regional network representation: nodes are banks or the market proxy and arrows follow the estimated coefficient orientation. The market proxy is first in the full-sample order. Its inclusion allows observed market variation to enter the representation; it does not make market shocks externally identified interventions.

\figone{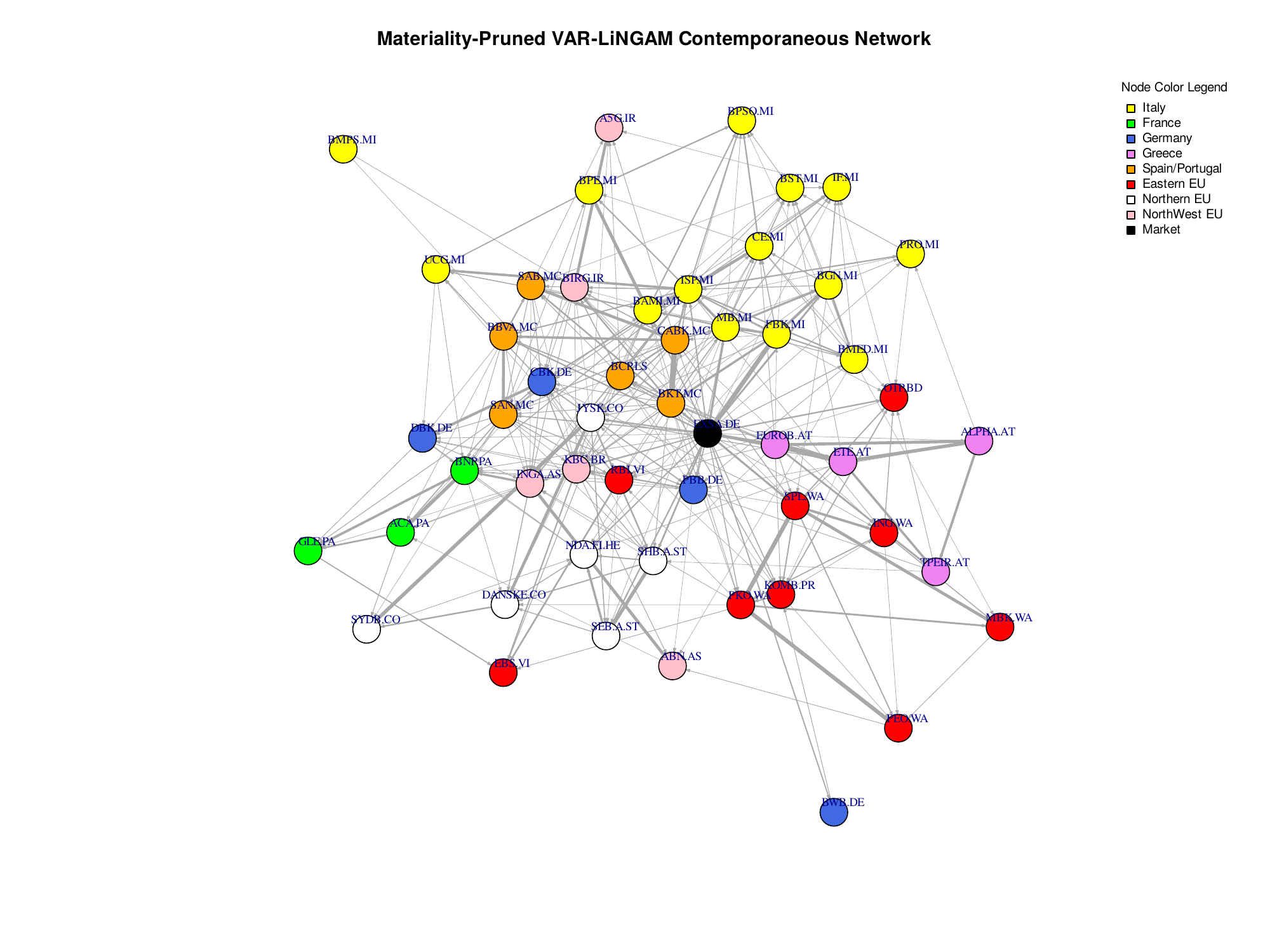}{Magnitude-screened contemporaneous network. The graph displays the 312 coefficients with absolute value at least 0.05 in the unpruned, volatility-standardized fit. Regional node colors are descriptive. Arrow orientation is conditional on the recursive model, and this dense graph is an overview rather than a collection of independently validated causal links.}{fig:full_network}{0.98}

The market's early position is compatible with broad common variation influencing bank returns. Individual banks can still receive and transmit effects through their estimated connections. The estimated network is sufficiently dense that no individual arrow should be treated as the principal finding. More informative is the heterogeneity in direct and indirect reach across banks and the distinction between a descriptive magnitude-pruned overview and the narrower network that survives the full-family coefficient-selection rule. The dense overview should be interpreted together with coefficient selection and discovery uncertainty rather than read arrow by arrow.

\subsection{Significance test}
\label{sec:wald_res}
\figone{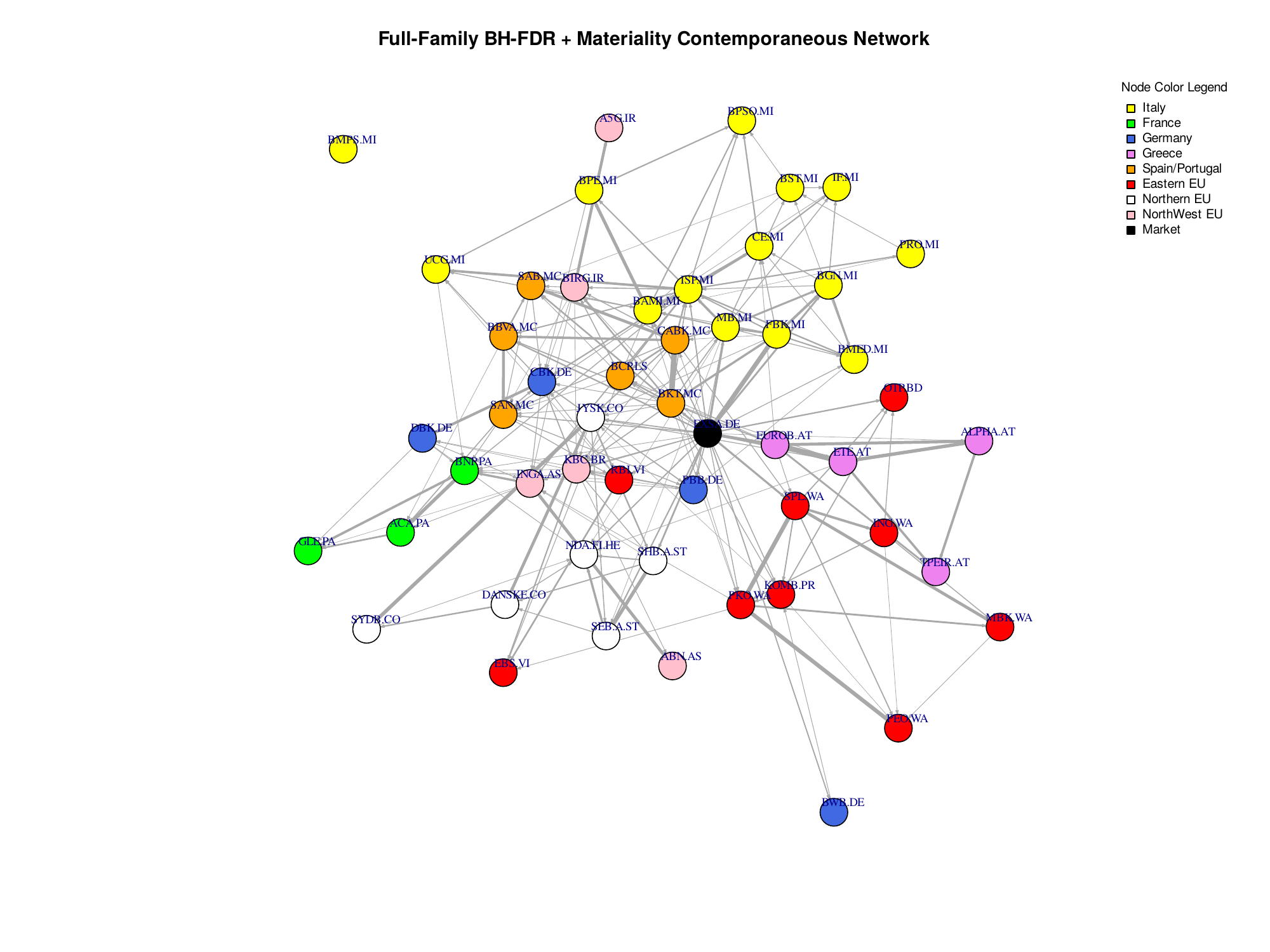}{Conditional Wald--BH network used for systemic scores. The 214 displayed links survive adjustment over all order-allowed coefficients and the subsequent 0.05 materiality rule. Inference is conditional on the selected order. Node positions and regional colors match Figure~\ref{fig:full_network}.}{fig:fdr_network}{0.98}

The conditional coefficient bootstrap produces 332 unadjusted $p$-values below 0.05 among the 1,326 allowed coefficients. Of these, 219 survive BH adjustment, and \Nedges{} also exceed the absolute-coefficient threshold. BY retains 170 coefficients before and 169 after materiality screening. Figure~\ref{fig:fdr_network} shows the BH-plus-materiality graph used for systemic-score calculations. Table~\ref{tab:networks} compares selection on filtered and unfiltered inputs using the same calendar-screened sample. The unfiltered input yields a network of similar density but substantially more independence rejections (Section~\ref{sec:independence_results}). Coefficient selection conditional on the recovered order does not assess the identifying independence assumption.

\begin{table}[!htbp]
\centering
\caption{Conditional coefficient selection on the same calendar-screened sample.}\label{tab:networks}
\small
\begin{tabular}{lrrrrr}
\toprule
Input & T & $|B|\ge .05$ & Unadjusted $p$ & BH & BH + materiality \\
\midrule
Volatility-standardized & 1990 & 312 & 332 & 219 & 214 \\
Unfiltered & 1990 & 351 & 324 & 193 & 188 \\
\bottomrule
\end{tabular}
\notetext{Unadjusted and BH columns count coefficient tests below 0.05. BH + materiality adds the 0.05 absolute-coefficient restriction after correction. The full family contains 1,326 order-allowed coefficients. Both specifications use 1,000 block resamples of length ten.}
\end{table}

The reduction from the magnitude-screened graph indicates that some seemingly material coefficients remain uncertain even when the order is held fixed. This is why applying significance screens after selecting only coefficients larger than the display threshold would answer a different, potentially more permissive multiple-testing question. The order-conditional exercise assesses sampling variation of coefficients among the admissible predecessors, whereas a separate re-discovery exercise is needed to examine how much the predecessor set and the edge orientations themselves can change. It does not address the separate possibility that a different order would assign another direction to a relation.

\subsection{Causal ordering stability}
\label{sec:stability_results}
\figone{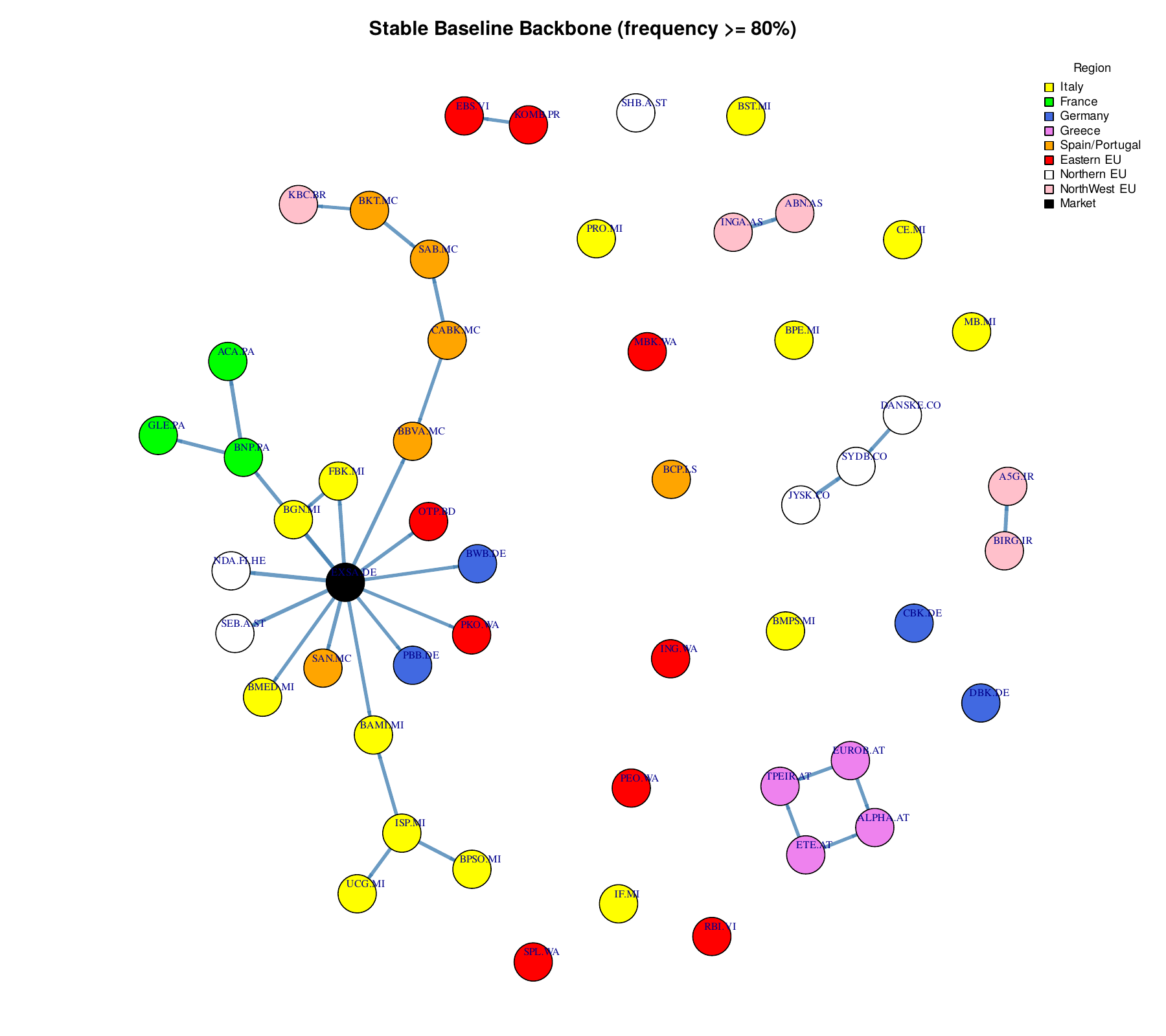}{Strict discovery backbone. These 32 baseline links recur in the same direction in at least 80\% of 500 re-discovery replications. Each replication repeats ordering, coefficient selection, and score calculation. Direction frequency refers to resampling reproducibility conditional on the fitted filter, not to verification of independent shocks.}{fig:backbone}{0.86}

Re-discovery is more demanding than re-estimating coefficients on a fixed graph. The 500 outer replications yield 624 unordered pairs active in at least 10\% of resamples under the summed-frequency definition. Of these, 17.8\% are directionally ambiguous. The strict backbone contains \Nbackbone{} baseline links whose directions recur in at least 80\% of replications. Of these, 19 are bank-to-bank links and 13 originate from the broad market proxy. Figure~\ref{fig:backbone} and Table~\ref{tab:backbone} describe this subset; Appendix Table~\ref{tab:backbone_full} lists all of its edges. The full frequency matrix appears in Appendix Figure~\ref{fig:frequency}.

\begin{table}[!htbp]
\centering
\caption{Largest coefficients in the strict discovery backbone.}\label{tab:backbone}
\small
\begin{tabular}{llrr}
\toprule
Source & Target & Coefficient & Frequency \\
\midrule
EXSA.DE & FBK.MI & 0.550 & 0.93 \\
JYSK.CO & SYDB.CO & 0.446 & 0.96 \\
BNP.PA & ACA.PA & 0.421 & 0.96 \\
INGA.AS & ABN.AS & 0.382 & 0.99 \\
ETE.AT & ALPHA.AT & 0.428 & 0.84 \\
BIRG.IR & A5G.IR & 0.350 & 0.93 \\
CABK.MC & SAB.MC & 0.360 & 0.89 \\
EUROB.AT & ALPHA.AT & 0.377 & 0.85 \\
FBK.MI & BGN.MI & 0.315 & 0.85 \\
BNP.PA & GLE.PA & 0.300 & 0.90 \\
ISP.MI & UCG.MI & 0.302 & 0.86 \\
CABK.MC & BBVA.MC & 0.302 & 0.82 \\
ETE.AT & TPEIR.AT & 0.287 & 0.83 \\
ISP.MI & BAMI.MI & 0.256 & 0.86 \\
EXSA.DE & BGN.MI & 0.229 & 0.95 \\
EUROB.AT & TPEIR.AT & 0.225 & 0.86 \\
\bottomrule
\end{tabular}
\notetext{The complete backbone contains 32 baseline edges with direction frequency at least 0.80 across 500 re-discovery replications. The displayed subset is ordered by absolute coefficient times directional frequency. Appendix Table~\ref{tab:backbone_full} lists all edges. Frequencies condition on the fitted volatility filter.}
\end{table}

\figone{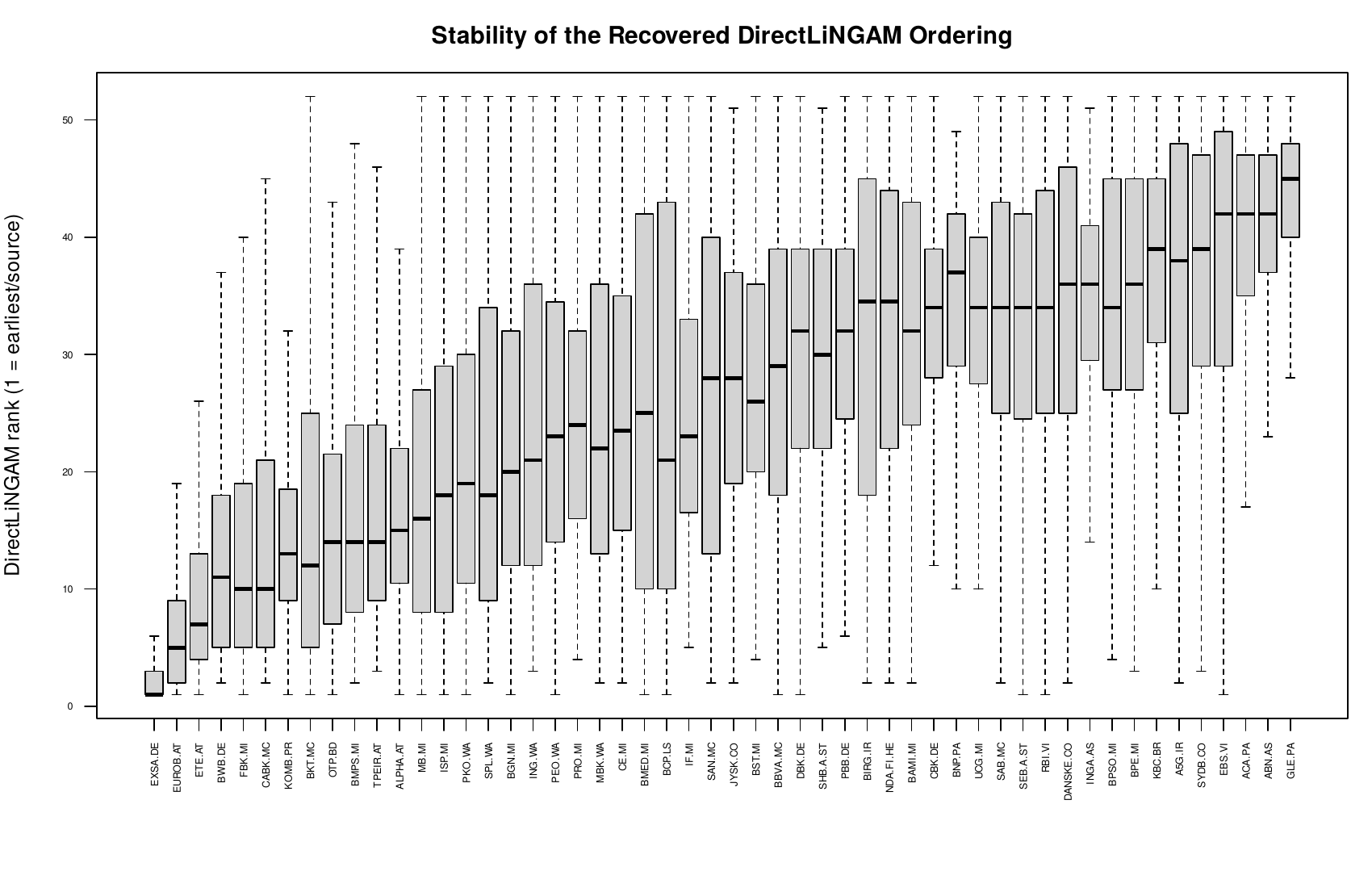}{Distribution of topological positions under re-discovery. Variables are ordered by mean position; boxes show interquartile ranges, central lines are medians, and whiskers extend to 1.5 interquartile ranges. Position 1 is earliest in the recursive representation.}{fig:order}{0.99}

Topological positions vary substantially across replications (Figure~\ref{fig:order}). An early position in one estimated order should therefore not be treated as a fixed property of a bank. More importantly, the median rank correlation between baseline and re-discovered scores is \BootSIS{} for SIS and \BootSVS{} for SVS; their fifth percentiles are 0.31 and 0.58. Receiver rankings are relatively more reproducible, but neither result establishes exact ranking invariance. This asymmetry matters for economic interpretation. A high full-sample impact rank identifies a candidate source under the fitted ordering; the wide re-discovery range is a reason to emphasize membership in a broader transmitting group over a precise ordinal position. Receiving roles are more reproducible in these samples, yet their intervals also preclude a claim that the hierarchy is immutable.

\FloatBarrier

\subsection{A diagnostic check of recursiveness}
\label{sec:recursiveness}
DirectLiNGAM imposes an acyclic contemporaneous structure. We assess it against an estimator that does not impose it. Let $\bm{u}_t^{*}$ collect the standardized baseline observations ordered as in the recovered DirectLiNGAM order, with sample covariance matrix $\bm{\Sigma}^{*}$ and lower-triangular Cholesky factor $\bm{D}$, so that $\bm{D}\bm{D}'=\bm{\Sigma}^{*}$. If the recursive representation holds in this order with mutually independent shocks, the mixing matrix in $\bm{u}_t^{*}=\bm{C}^{*}\bm{\varepsilon}_t^{*}$ is lower triangular and, with shocks scaled to unit variance, equals $\bm{D}$ in population. An independent component analysis (ICA) estimate $\widetilde{\bm{C}}$ of $\bm{C}^{*}$ should then be close to $\bm{D}$, up to permutations and scale changes of its columns, since ICA identifies components but not their order, sign, and variance. FastICA fixes the scale by giving the components unit variance, so that $\widetilde{\bm{C}}\widetilde{\bm{C}}'=\bm{\Sigma}^{*}$. We resolve order and sign by the closest signed permutation of the columns of $\widetilde{\bm{C}}$:
\begin{equation}
 \mathcal{D}(\widetilde{\bm{C}},\bm{D})=\min_{\bm{P}\in\mathcal{P}_{\pm}}\big\|\widetilde{\bm{C}}\bm{P}-\bm{D}\big\|_F,
 \label{eq:recursive_distance}
\end{equation}

where $\mathcal{P}_{\pm}$ is the set of signed permutation matrices. This is closely related to the minimum-distance index of \citet{ilmonen2010new}, which, differently from (\ref{eq:recursive_distance}),
is scale-invariant and 
lies between zero and one and is zero when $\widetilde{\bm{C}}^{-1}\bm{D}$ is a rescaled signed permutation. The reference distribution replaces $\widetilde{\bm{C}}$ by $\widetilde{\bm{C}}\bm{R}^{(i)}$ for 10,000 uniformly random rotation matrices $\bm{R}^{(i)}$. These are alternative factorizations of the same covariance matrix, and each is aligned in the same way. A second reference rotates the Cholesky factor instead and computes $\mathcal{D}(\widetilde{\bm{C}},\bm{D}\bm{R}^{(i)})$ with the same rotations. Appendix~\ref{app:recursiveness} gives the computational details.

The aligned distance is 4.18, against a median of 7.08 and a minimum of 6.60 across the rotated ICA estimates, and a median of 6.57 and a minimum of 6.35 across the rotated Cholesky factors. The recursive factor is therefore closer to the ICA estimate than all 10,000 rotated factorizations under either reference. The minimum-distance index is 0.38, against a median of 0.95 for the rotations under either reference. All four comparisons give the same result for ten FastICA starting values and for either covariance divisor. Without the alignment, the comparison is not informative: the share of rotated factorizations farther from the ICA estimate than $\bm{D}$ ranges from 2\% to 95\% across FastICA starting values, because it reflects the arbitrary order and signs of the estimated components. The rotation reference shows that the recursive factorization in the recovered order is far closer to the independent-component representation than an arbitrary factorization of the same covariance matrix. It does not show that the representation is exactly recursive. In 20 exactly acyclic systems simulated from the selected coefficient matrix, the aligned distance has median 1.87 (5th--95th percentile 1.80--1.98) and the minimum-distance index 0.26. The empirical distance is more than twice this finite-sample reference. The backward-mass diagnostic in Appendix~\ref{app:recursiveness} leads to the same reading. A contemporaneous recursive ordering in daily data is a representation of within-day relationships after aggregation, not a direct measurement of the sequence of every transaction or information arrival within each day. This supports treating recursiveness as an approximation. It does not verify absence of intraday feedback or latent confounding.

\begin{table}[!htbp]
\centering
\caption{Exact Gaussian-kernel pairwise independence diagnostics.}\label{tab:hsic}
\small
\begin{tabular}{lrrrrr}
\toprule
Specification & Unadjusted & BH & BY & Median $p$ & Sq. corr. \\
\midrule
Baseline & 219 & 14 & 0 & 0.312 & 0.020 \\
Unfiltered & 481 & 276 & 0 & 0.115 & 0.045 \\
Return-magnitude screen: 0.35 & 216 & 14 & 0 & 0.310 & 0.021 \\
Training window & 107 & 0 & 0 & 0.378 & 0.016 \\
VAR(1) & 190 & 0 & 0 & 0.354 & 0.019 \\
\bottomrule
\end{tabular}
\notetext{Unadjusted, BH, and BY count rejections at 5\% among 1,326 recovered-shock pairs. Sq. corr. is the mean pairwise correlation of squared shocks. The primary 1,990-observation sample has 1,951 admissible circular shifts at minimum circular distance 20, so the smallest attainable $p$-value is $1/1{,}952$; any BH rejection then involves at least 14 pairs and any BY rejection at least 106. Shocks come from the unpruned model. Non-rejection is not verification of joint independence.}
\end{table}

\FloatBarrier

\subsection{Independence test}
\label{sec:independence_results}

Table~\ref{tab:hsic} reports exact-kernel HSIC results for unpruned disturbances. In the baseline, 219 of 1,326 pairs reject before adjustment, 14 reject after BH, and none reject after BY. Thus, about 1.1\% of pairs retain a BH rejection. These 14 comparisons lie at the minimum attainable shift-reference $p$-value, $1/1{,}952$, and have BH-adjusted values of approximately 0.0485. The $p$-value floor imposes a resolution constraint: a non-empty BH rejection set must contain at least 14 comparisons at the designated level, and the corresponding lower bound under BY is 106. These are necessary constraints on the size of a rejection set, not sufficient conditions for rejection; the actual decision uses the complete ordered $p$-value family. The Simes combination statistic, numerically equal to the smallest BH-adjusted $p$-value, provides an additional summary of the family. Its interpretation as a calibrated global test depends on the relevant assumptions about the joint distribution of the $p$-values. For three pairs, Monte dei Paschi with the market proxy, SEB with Handelsbanken, and Banco Comercial Português with Monte dei Paschi, the observed statistic is 8.2, 3.3, and 2.3 times the largest shifted statistic. Median normalized HSIC is 0.00166 and the mean correlation of squared shocks is 0.020. On the same observations without volatility filtering, 276 pairs, or 20.8\%, reject after BH and squared-shock correlation averages 0.045.

Volatility standardization substantially reduces detectable dependence in the recovered shocks. Relative to the unfiltered specification, the number of BH-rejected pairs decreases from 276 to 14, and the mean pairwise correlation of squared shocks decreases from 0.045 to 0.020. These changes provide positive diagnostic evidence that filtering reduces detectable dependence on the fitted scale. Nevertheless, the full-sample baseline does not qualify for pairwise independence support under the designated 5\% BH classifier, following the operational reading in \citet[Section 4.3]{doannguyenbucciol2026}. The classification is close to the discrete $p$-value boundary; the more conservative BY correction yields no rejections, and aggregate dependence-severity measures are low, although several individual pairs show appreciable departures relative to the shift reference. These qualifications describe the strength and sensitivity of the diagnostic evidence but do not reverse the designated BH decision. Accordingly, SIS and SVS are reported as model-implied contemporaneous transmission sensitivities. A structural interpretation remains conditional on the maintained identifying assumptions, which are not independently confirmed by the baseline diagnostic.

By contrast, the pre-2022 training-window and VAR(1) specifications yield no BH rejections at the designated 5\% level. These fitted representations therefore have pairwise independence support in the operational sense of \citet[Section 4.3]{doannguyenbucciol2026}.\footnote{The shorter training window has a coarser circular-shift reference. With 1,267 observations and minimum circular distance 20, there are 1,228 admissible shifts and the $p$-value floor is $1/1{,}229$. At BH 5\% over 1,326 pairs, a non-empty rejection set must contain at least 22 comparisons. Consequently, non-rejection in this window should not be ranked as stronger evidence than a full-sample verdict solely by comparing rejection counts.} This is positive, specification-specific diagnostic evidence, not proof of joint mutual independence and not transferable to the full-sample baseline. Neither result by itself establishes the remaining conditions for a structural or independently supported directional interpretation. Each recovered network must therefore be evaluated under its own identifying diagnostics and maintained assumptions. The return-magnitude-screened specification also yields 14 BH rejections, with eight rejected pairs overlapping with the baseline. Thus, the aggregate rejection count is unchanged, although pair-level classifications differ.

\subsection{Tail co-exceedance and temporal holdout diagnostics}
\label{sec:tail_results}
The selected network groups pairs with stronger lower-tail dependence. The symmetric and shock-conditioned statistics address distinct questions: excess tail association among linked banks concerns the economic relevance of selected linkages, while directional conditioning is sensitive to the fitted unmixing representation and does not identify causality merely through an asymmetry. Mean CoLift is 7.39 for linked pairs and 5.32 for genuine non-links, a ratio of 1.39. The shock-conditioned lift is 3.45 in the estimated direction, compared with 1.74 in reverse and 1.89 for order-respecting non-links (Figures~\ref{fig:tail_coex} and \ref{fig:tail_directed}). CoLift itself is symmetric and does not identify arrow orientation.
\figone{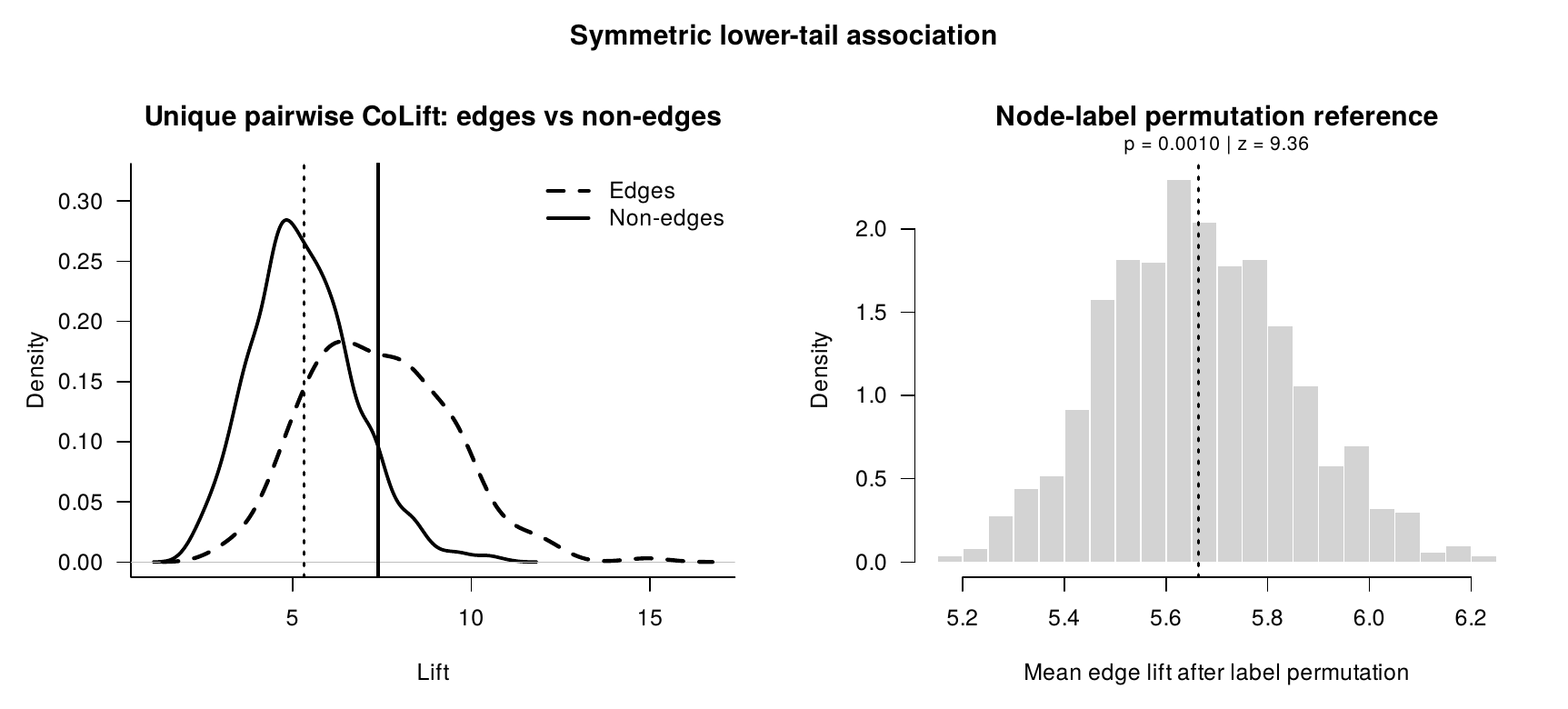}{Symmetric lower-tail co-exceedance in the full sample at the 5\% threshold: linked versus genuinely unlinked pairs and the node-label permutation reference. The statistic is symmetric, not an orientation test.}{fig:tail_coex}{0.88}

\begin{table}[!htbp]
\centering
\caption{Symmetric and shock-conditioned tail diagnostics.}\label{tab:tail}
\footnotesize
\begin{tabular}{lrrrrrr}
\toprule
Sample & Edges & CoLift ratio & Forward & Reverse & Non-link & Gap \\
\midrule
Full-sample & 214 & 1.39 & 3.45 & 1.74 & 1.89 & 1.71 \\
Temporal holdout & 150 & 1.41 & 4.39 & 2.37 & 2.51 & 2.02 \\
Holdout, bank edges & 130 & 1.45 & 3.92 & 2.41 & 2.38 & 1.51 \\
Holdout, fewer than 10 & 150 & 2.90 & 3.45 & 1.62 & 1.44 & 1.83 \\
Holdout, fewer than 20 & 150 & 2.11 & 3.82 & 1.77 & 1.76 & 2.05 \\
\bottomrule
\end{tabular}
\notetext{CoLift ratio divides the mean over linked pairs by the mean over genuine non-links. The holdout uses 1,267 training and 723 test observations. Return and shock thresholds are fixed from training. The last two rows remove 45 and 14 test sessions, respectively, based on the number of bank co-exceedances. Most rows include the market node; the bank-edge row excludes its incident edges. A positive gap is not independent evidence of causal direction.}
\end{table}

\figone{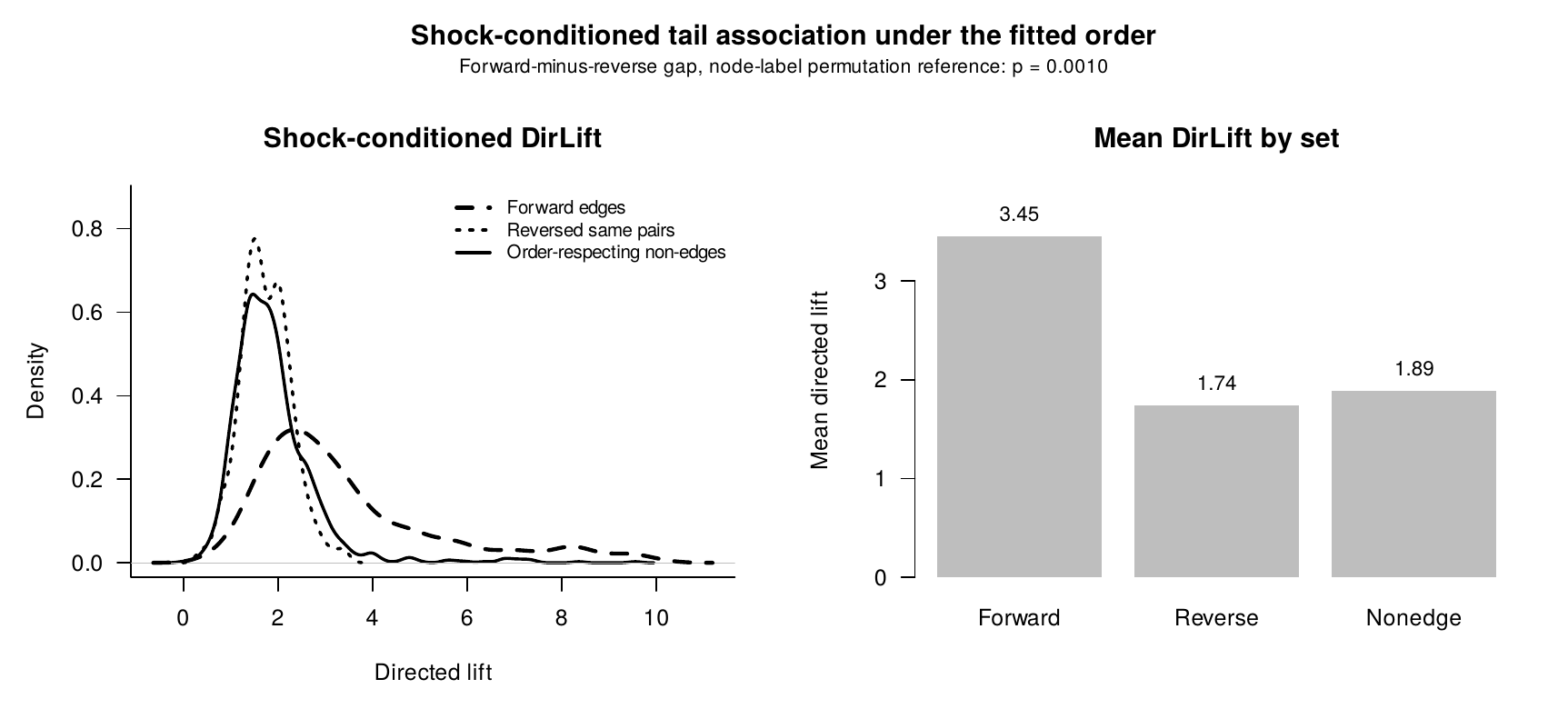}{Full-sample shock-conditioned tail lift in the estimated direction, the reverse direction for the same pairs, and genuine non-links. These same-period quantities depend on the recovered shock transformation.}{fig:tail_directed}{0.91}

\FloatBarrier

For the temporal holdout, the model, volatility-filter parameters and tail thresholds are estimated on 1,267 observations through 2021 and evaluated on 723 later observations. Without re-estimating the fitted transformation in the test period, forward and reverse lifts are 4.39 and 2.37, respectively, a gap of 2.02 (Table~\ref{tab:tail}); the ten-observation block-resampling interval for the gap is [1.63, 2.56]. The corresponding bank-to-bank gap is 1.51. The separation between the market-inclusive and bank-only results matters for a systemic-risk application: part of the system-wide association may be organized around the observed market node rather than around a bilateral bank-to-bank mechanism. Excluding dates with widespread lower-tail co-exceedances leaves a positive gap. Out of sample, the symmetric co-exceedance ratio of linked to unlinked pairs is 1.41 (1.39 in sample) and rises to 2.90 and 2.11 when the sessions with ten or more and twenty or more bank co-exceedances are removed (Table~\ref{tab:tail}); this part of the holdout evidence does not depend on orientation. The exercise examines the transport of a same-day conditional dependence pattern, not a forecast of future returns or a test of interventions.

The holdout results show a positive forward--reverse tail asymmetry under the fitted pre-2022 representation. However, 42\% of coherently re-estimated alternative recursive orders produce a gap at least as large as that of DirectLiNGAM, so the comparison does not independently establish the recovered orientation. Relative to the directional-standing standard of \citet[Definition 3.4]{doannguyenbucciol2026}, the evidence is insufficient to establish criterion (i); the benchmark comparisons document distinctness but do not yet establish incremental information under criterion (ii); and robustness to a genuinely alternative recovery algorithm under criterion (iii) has not been assessed. These qualifications concern the interpretation of orientation, rather than eliminating the observed out-of-sample tail associations of the fitted network. These alternative orders are not necessarily independence-compatible LiNGAM models; the comparison is a diagnostic, not an exact causal-null test.

\begin{table}[!htbp]
\centering
\caption{Holdout gap under coherently re-estimated alternative orders.}\label{tab:coherent}
\small
\begin{tabular}{p{8.1cm}rr}
\toprule
Specification & Selected edges & Forward--reverse gap \\
\midrule
DirectLiNGAM order & 150 & 2.023 \\
Alternative orders: mean & 171.6 & 1.979 \\
Alternative orders: median & 171.0 & 1.981 \\
Alternative orders: 5th--95th percentile & [157, 189] & [1.635, 2.271] \\
\bottomrule
\end{tabular}
\notetext{For 200 alternative orders, the market is kept first and banks are randomly reordered. Each order has its own fitted predecessor coefficients, selected network, recovered shocks, and shock thresholds. 42.0\% of these fits have a gap at least as large as the DirectLiNGAM fit. The comparison uses 1,000 inner coefficient resamples for every alternative and for the baseline; selected graph densities differ. This is a diagnostic comparison, not an exact causal-null test.}
\end{table}

\FloatBarrier

\subsection{Systemic impact and vulnerability indices}
\label{sec:indices_results}
\subsubsection{Total contemporaneous transmission}

Figure~\ref{fig:total} displays the bank-only total-effect matrix obtained from the selected network. The self-effect diagonal is omitted visually so that cross-bank transmission is readable. Effects are accumulated along directed paths before their absolute values enter the scores. Negative total effects account for 0.4\% of aggregate absolute SIS, so the reported ranking is predominantly driven by positive rather than offsetting net responses. A column of the total-effect matrix describes the set of banks whose modeled returns respond directly or through intermediate paths to a disturbance at the originating bank. The associated row instead describes the institution's aggregate exposure to disturbances originating elsewhere. These complementary views motivate the separate aggregation of source and receiver sensitivities rather than treating graph degree or bank size as a sufficient description of systemic position.

\figone{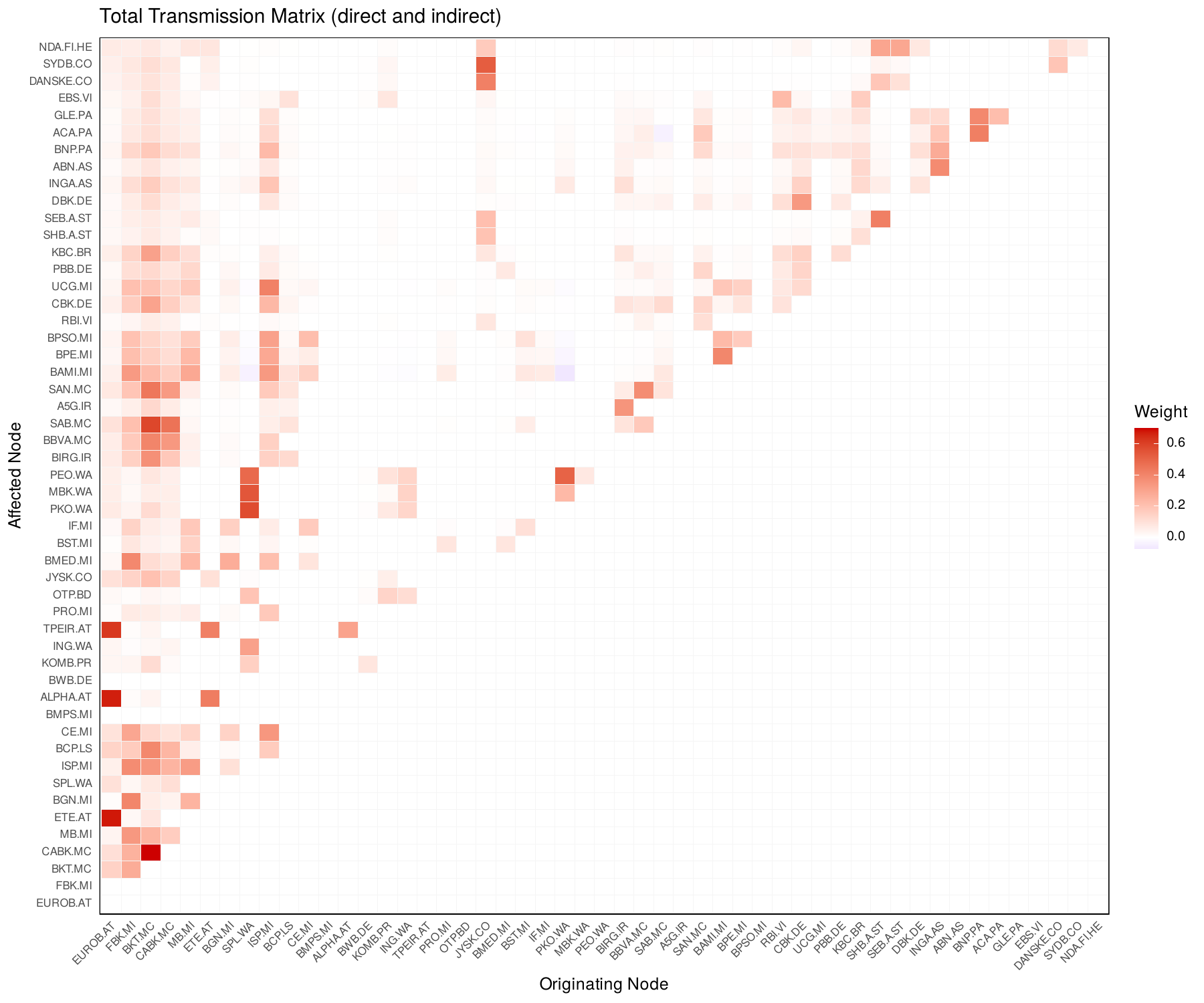}{Bank-only total contemporaneous effects. Sources are columns and recipients are rows, arranged in baseline order after removing the market node. The unit diagonal is set to zero only for this display; off-diagonal entries are unchanged. Scores exclude own effects and use the same matrix.}{fig:total}{0.94}

\FloatBarrier

\subsubsection{Source and receiver rankings}
SIS and SVS identify different roles (Table~\ref{tab:rankings} and Figure~\ref{fig:sis_svs}). The rank correlation between the two scores is $-0.02$. Bankinter, FinecoBank, CaixaBank, Intesa Sanpaolo, and Mediobanca form the five highest full-sample SIS observations. Cr\'edit Agricole, Soci\'et\'e G\'en\'erale, BNP Paribas, Santander, and UniCredit are the five highest SVS observations. These are conditional full-sample rankings, not time-invariant classifications. The distinction is economically important: the largest institution need not be the leading source of contemporaneous return sensitivity, while an institution central to the absorption of disturbances can differ substantially from those early in the model-implied ordering. The SIS/SVS quadrant representation makes this source--receiver asymmetry transparent without interpreting the categories as observed balance-sheet transmission roles.

Impact is more concentrated: the top five institutions account for \TopFiveSIS\% of aggregate SIS, compared with \TopFiveSVS\% of aggregate SVS. Rank correlations with capitalization are \SISsize{} for impact and \SVSsize{} for vulnerability. Setting capitalization weights equal to one produces correlations of 0.95 and 0.89 with the weighted SIS and SVS rankings. Weighting affects the object being measured, but does not alone generate the distinction between source and receiver roles. In SIS the capitalization weight belongs to the affected institution, whereas in SVS it belongs to the originating institution. Because the location of the weight differs, common associations with size should be measured rather than assumed away. The size-neutral sensitivity helps establish whether the empirical distinction depends predominantly on the 1 July 2025 capitalization snapshot.

\begin{table}[!htbp]
\centering
\caption{Leading banks and re-discovery rank uncertainty.}\label{tab:rankings}
\small
\begin{tabular}{lrrrrr}
\toprule
Ticker & Score & Rank & Median rank & 5--95\% range & Top-five freq. \\
\midrule
\multicolumn{6}{l}{\textit{Systemic impact}} \\
BKT.MC & 269.5 & 1 & 5.0 & [1, 31] & 0.56 \\
FBK.MI & 175.8 & 2 & 10.0 & [2, 37] & 0.35 \\
CABK.MC & 160.3 & 3 & 4.0 & [1, 27] & 0.65 \\
ISP.MI & 139.5 & 4 & 5.0 & [1, 25] & 0.57 \\
MB.MI & 98.8 & 5 & 9.0 & [1, 40] & 0.35 \\
EUROB.AT & 68.0 & 6 & 13.0 & [4, 38] & 0.12 \\
CBK.DE & 57.4 & 7 & 17.5 & [3, 42] & 0.13 \\
BBVA.MC & 56.6 & 8 & 11.0 & [3, 49] & 0.22 \\
\multicolumn{6}{l}{\textit{Systemic vulnerability}} \\
ACA.PA & 96.4 & 1 & 2.0 & [1, 11] & 0.83 \\
GLE.PA & 91.1 & 2 & 2.0 & [1, 13] & 0.77 \\
BNP.PA & 82.7 & 3 & 6.0 & [2, 46] & 0.45 \\
SAN.MC & 71.3 & 4 & 9.0 & [3, 48] & 0.24 \\
UCG.MI & 63.3 & 5 & 7.0 & [3, 21] & 0.29 \\
CBK.DE & 60.1 & 6 & 7.0 & [2, 25] & 0.37 \\
BAMI.MI & 54.2 & 7 & 7.0 & [1, 20] & 0.42 \\
SAB.MC & 53.4 & 8 & 6.0 & [1, 25] & 0.44 \\
\bottomrule
\end{tabular}
\notetext{Rank 1 is highest. Percentile ranges summarize 500 re-discovery fits and are rounded to whole ranks; medians can be fractional because ranks are averaged for ties. These are descriptive bootstrap distributions, not simultaneous confidence intervals. GARCH filtering is held fixed during these replications.}
\end{table}

\figone{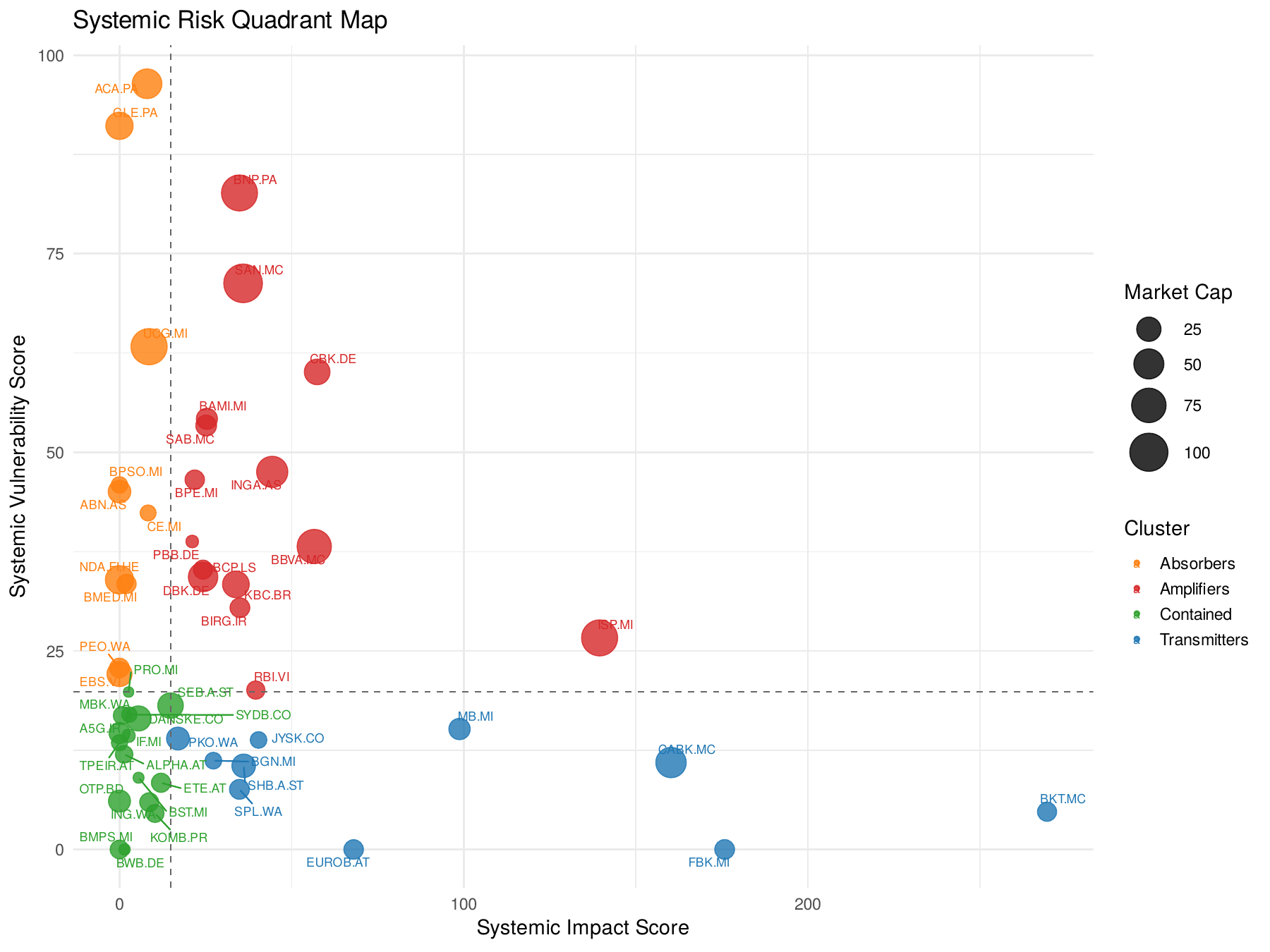}{Systemic impact versus systemic vulnerability. Marker size increases with market capitalization, and dashed lines are the sample medians. All banks are labeled and grouped descriptively by the two median thresholds. Scores are sensitivities on the filtered-return scale, not expected losses.}{fig:sis_svs}{0.90}

\subsubsection{Sensitivity of bank roles to volatility filtering}
\label{sec:matched}
\begin{table}[!htbp]
\centering
\caption{Unfiltered and volatility-standardized bank roles on the same 1,990 observations.}\label{tab:matched}
\small
\begin{tabular}{lccc}
\toprule
Score & Rank correlation & Top-five overlap & Median absolute rank change \\
\midrule
SIS & \MatchedSISRho & \MatchedSISTopFive/5 & \MatchedSISMedianRankShift \\
SVS & \MatchedSVSRho & \MatchedSVSTopFive/5 & \MatchedSVSMedianRankShift \\
\bottomrule
\end{tabular}
\notetext{Spearman correlation of the 51 bank-level scores across the two fits; rank 1 is highest and tied scores receive average ranks. Both fits re-estimate the ordering and apply the same Wald--BH and materiality selection (Table~\ref{tab:networks}). Top five without volatility filtering: SIS BKT.MC, BNP.PA, FBK.MI, ISP.MI, BIRG.IR; SVS SAN.MC, BAMI.MI, ACA.PA, UCG.MI, BPE.MI. The volatility-standardized top five are those of Table~\ref{tab:rankings}.}
\end{table}

To separate the input transformation from changes in sample composition, we compare unfiltered and volatility-standardized fits on the same 1,990 retained observations. Both specifications use the same bank universe, capitalization weights and coefficient-selection procedure, with each ordering re-estimated. Tables~\ref{tab:networks} and~\ref{tab:hsic} report the associated selection and independence diagnostics. The matched comparison yields rank correlations of \MatchedSISRho{} for SIS and \MatchedSVSRho{} for SVS, with three common institutions in each top-five group (Table~\ref{tab:matched}). The median absolute rank changes are \MatchedSISMedianRankShift{} and \MatchedSVSMedianRankShift{}, respectively. This comparison quantifies the sensitivity of estimated bank roles to volatility filtering and the associated re-discovery on a fixed sample; it does not establish equivalence of the two structural representations. In both fits, Bankinter, FinecoBank and Intesa Sanpaolo are among the five highest SIS observations and Cr\'edit Agricole, Santander and UniCredit among the five highest SVS observations, whereas CaixaBank and Mediobanca (SIS) and Soci\'et\'e G\'en\'erale and BNP Paribas (SVS) enter the leading groups only after filtering; institution-level attribution beyond these banks therefore depends on the input transformation.

\subsubsection{Size-relative vulnerability}
Dividing vulnerability by own capitalization changes its interpretation and ranking. Banca Profilo, Banca Sistema, and Pfandbriefbank have the largest nSVS values; the rank correlation of nSVS with capitalization is $-0.47$. Figure~\ref{fig:sis_nsvs} retains the comparison with SIS. Because own size is in the denominator, a small-cap institution can have high nSVS without representing a large absolute system-wide loss. This score is accordingly reported as incoming sensitivity relative to size.

\figone{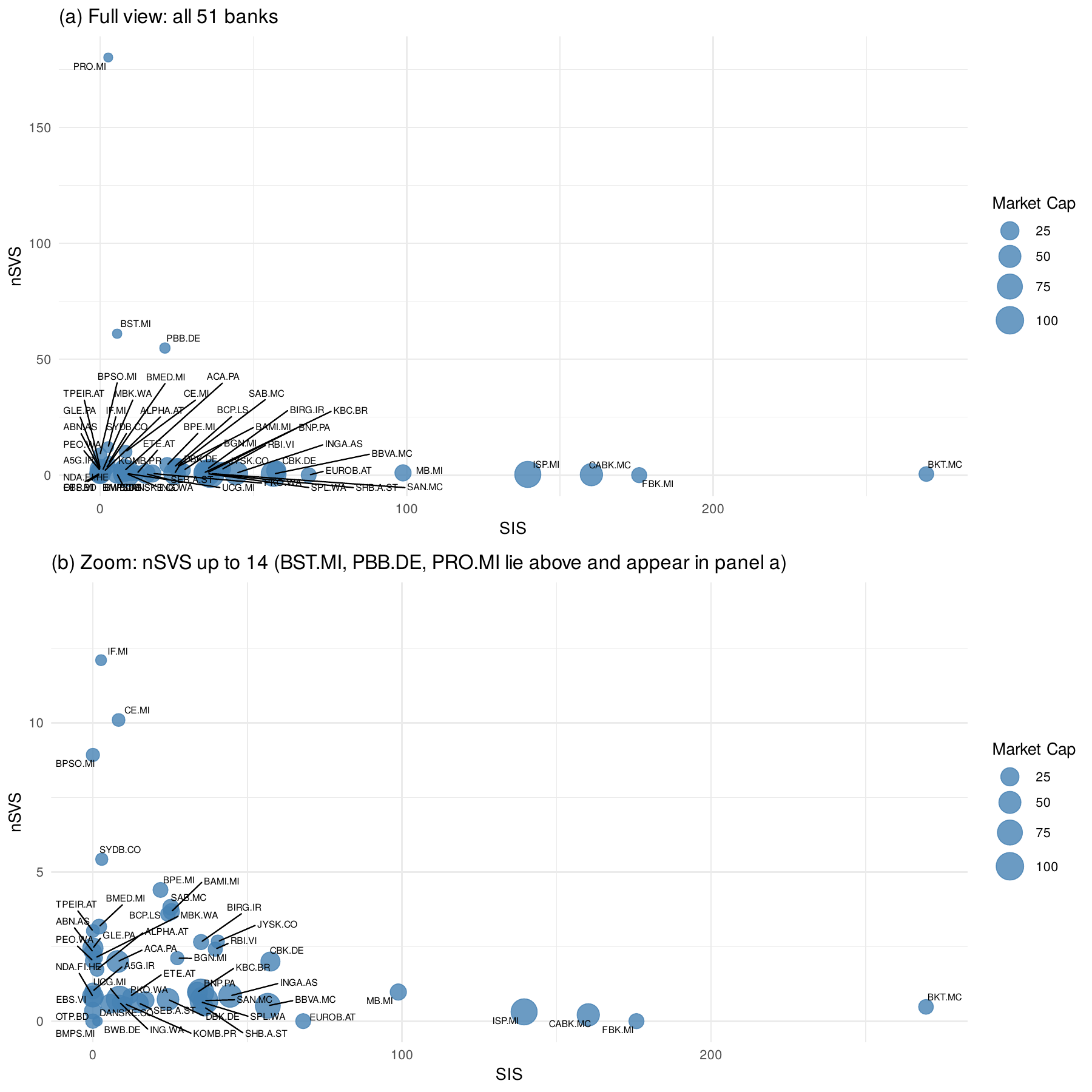}{Systemic impact and size-relative vulnerability, $\operatorname{nSVS}_i=\operatorname{SVS}_i/m_i$. Panel (a) shows all 51 banks; panel (b) shows the same scores with the vertical axis limited to nSVS values up to 14, which leaves out Banca Profilo, Banca Sistema and Pfandbriefbank (shown in panel a). Marker size increases with capitalization, and both axes are linear. The scale-relative ranking should not be interpreted as a ranking of absolute systemic losses.}{fig:sis_nsvs}{0.80}

\FloatBarrier

\subsubsection{Robustness to discovery uncertainty}
The uncertainty ranges in Table~\ref{tab:rankings} qualify the point rankings. CaixaBank and Intesa Sanpaolo are in the SIS top five in approximately 65\% and 57\% of replications, Bankinter in 56\%, FinecoBank in 35\%, and Mediobanca in 35\%. Exact source ranks are particularly uncertain. Cr\'edit Agricole and Soci\'et\'e G\'en\'erale appear in the SVS top five in approximately 83\% and 77\% of replications. Figure~\ref{fig:sis_uncertainty} makes the distinction between the full-sample rank and its re-discovery distribution explicit; Figure~\ref{fig:svs_uncertainty} reports the receiver counterpart.

\figone{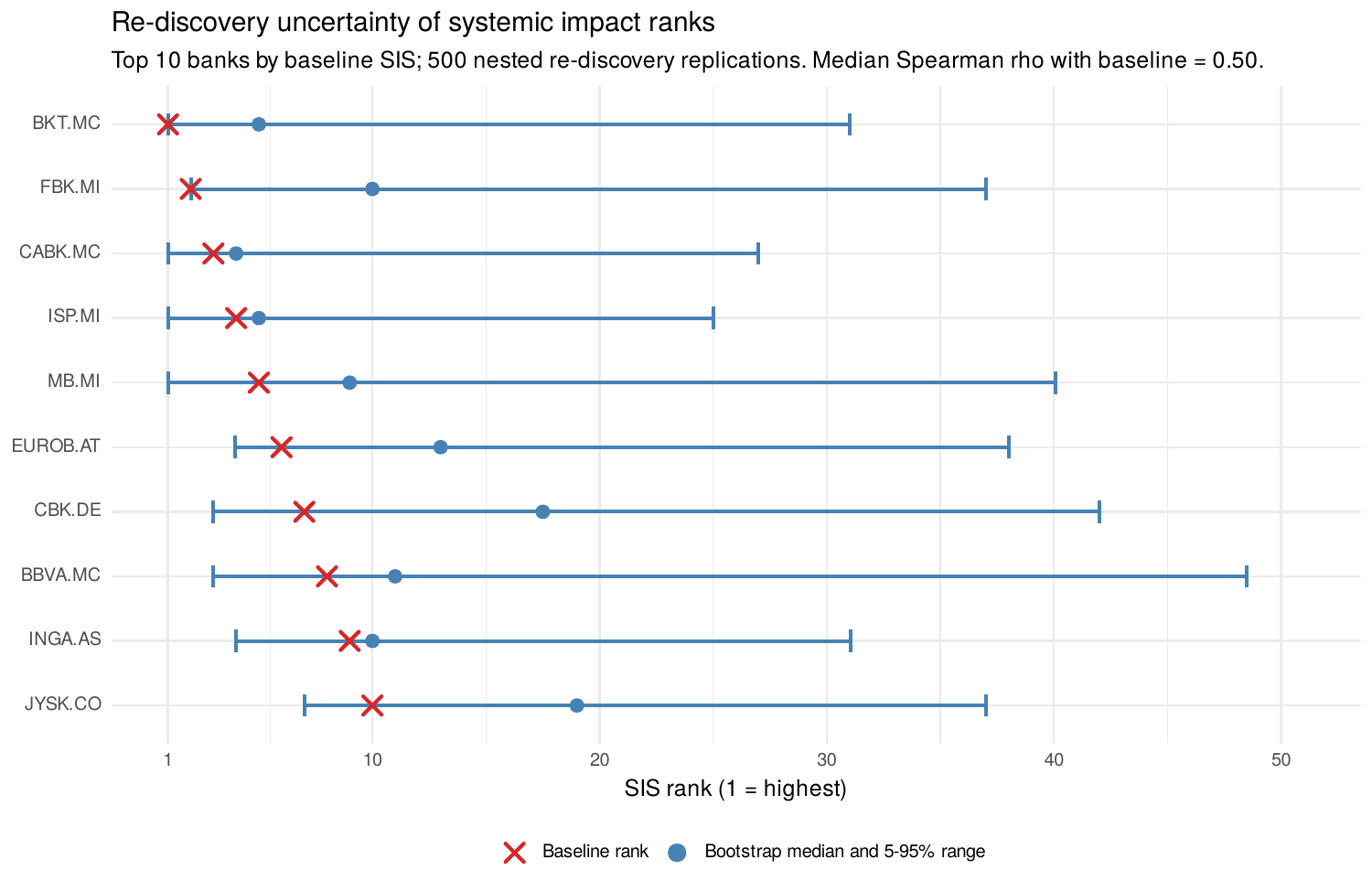}{Re-discovery uncertainty for the ten highest baseline impact scores. Crosses mark baseline ranks; circles and bars give bootstrap medians and 5th--95th percentile ranges. Rank 1 is highest. A high point rank does not imply that an institution occupies that rank across alternative recovered orders.}{fig:sis_uncertainty}{0.86}

\figone{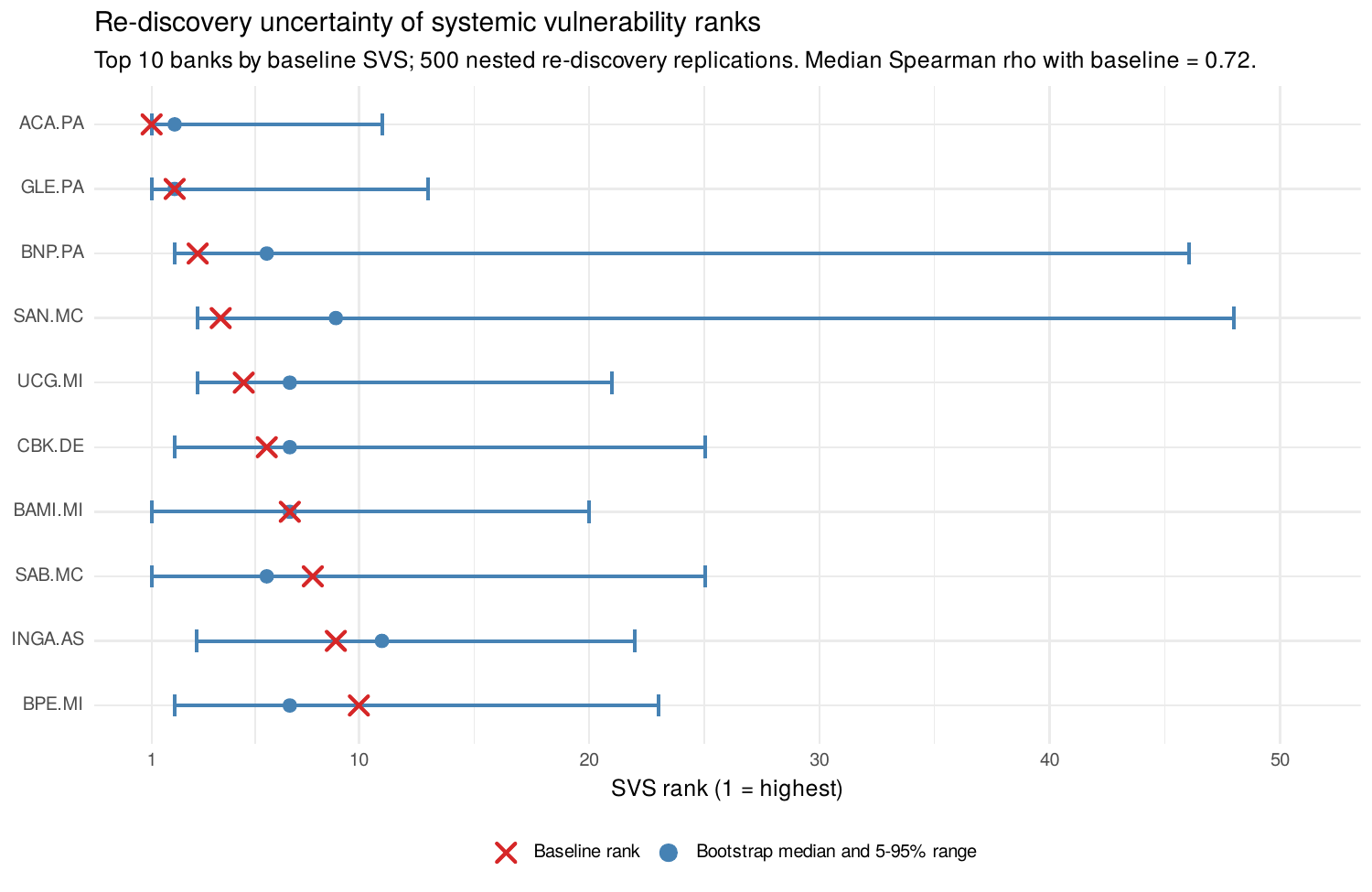}{Re-discovery rank uncertainty for the ten largest baseline vulnerability scores. Crosses are baseline ranks; circles and bars are bootstrap medians and 5th--95th percentile ranges. These distributions condition on the fitted volatility filter.}{fig:svs_uncertainty}{0.86}

\FloatBarrier

\subsection{Comparison with Diebold--Yilmaz connectedness and MES}
\label{sec:res_benchmarks}
Table~\ref{tab:benchmarks} reports score correlations with conventional connectedness and market-tail exposure. Total return-based DY connectedness is 86.7\%. SIS correlates 0.42 with ``to others'' connectedness, 0.30 after controlling for size, and approximately zero after also controlling for the first principal-component loading. The first component explains 42.6\% of standardized bank-return variation. SVS correlates 0.72 with ``from others'' connectedness, falling to 0.64 after size control and $-0.04$ after the additional common-loading control.

\begin{table}[!htbp]
\centering
\caption{Rank correlations with connectedness and market-tail exposure.}\label{tab:benchmarks}
\small
\begin{tabular}{llrrr}
\toprule
Score & Benchmark & Raw & Size control & Size + PC1 \\
\midrule
SIS & DY to & 0.416 & 0.302 & -0.006 \\
SVS & DY from & 0.721 & 0.639 & -0.043 \\
SIS & MES & 0.149 & 0.040 & -0.033 \\
SVS & MES & 0.449 & 0.335 & 0.252 \\
\bottomrule
\end{tabular}
\notetext{Partial rank correlations use residuals from linear regressions of ranks on ranked controls. DY uses a VAR(1) on standardized bank returns and a ten-observation generalized variance decomposition. MES uses raw bank returns on the worst 5\% of raw market-proxy returns. These comparisons do not test incremental predictive performance.}
\end{table}

\figone{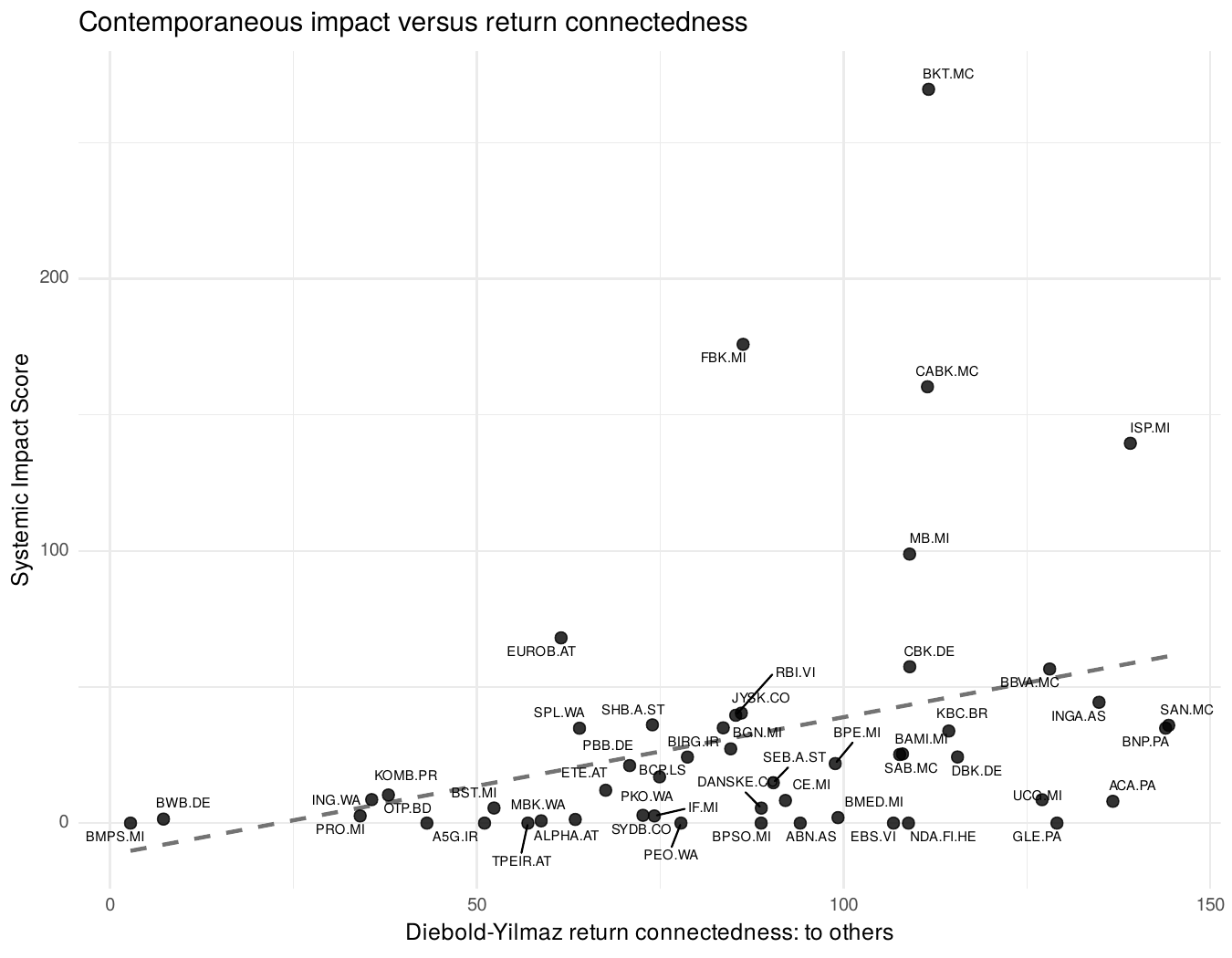}{Systemic impact versus generalized connectedness to others. The raw association is reduced after controlling for bank size and the first principal-component loading, as reported in Table~\ref{tab:benchmarks}.}{fig:bench_sis}{0.84}
\figone{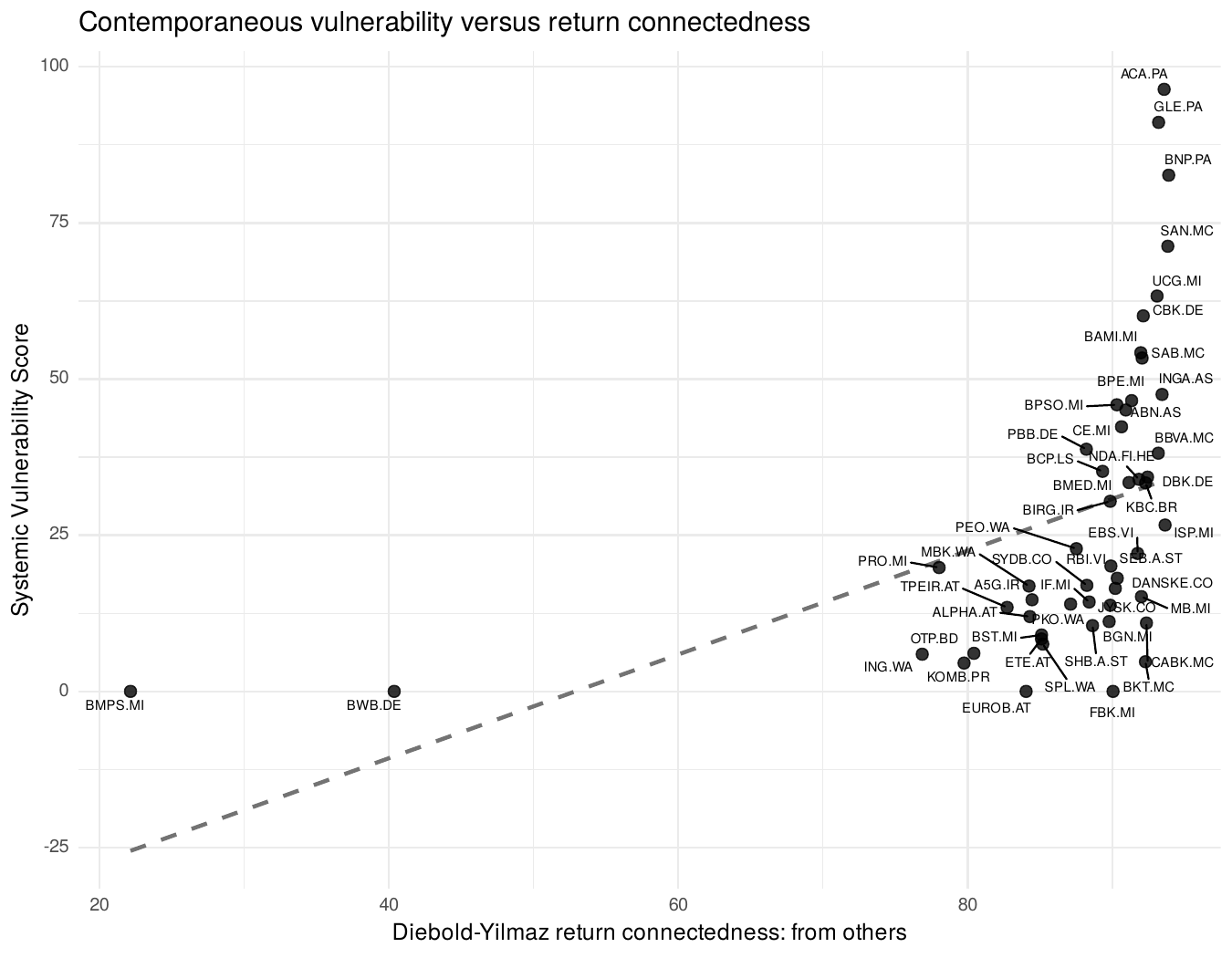}{Systemic vulnerability and generalized connectedness from others. This reduced-form benchmark summarizes forecast-error variance shares, while SVS summarizes weighted contemporaneous total effects under the recursive model.}{fig:bench_svs}{0.84}

\FloatBarrier

SIS has a weak association with MES (0.15), which falls close to zero under the same controls. SVS has a stronger MES correlation (0.45), retaining a partial correlation of 0.25 after both controls. Thus market-tail exposure overlaps more with receiving than with transmitting sensitivity in this sample. These differences establish neither causal validity nor incremental predictive value. Their economic significance lies in distinguishing estimands: generalized variance-decomposition connectedness allocates reduced-form forecast-error variation, MES conditions on market downturns, and SIS/SVS aggregate total same-period effects in the maintained recursive model. A disagreement among rankings is therefore not an estimation failure; it is an empirical question whether different concepts of systemic importance select the same institutions. In particular, disappearance of a correlation after a control does not by itself prove that the remaining score contains useful new information.

\subsection{Robustness to the return sample and lag specification}
\label{sec:var1_robustness}
Table~\ref{tab:robustness} reports the consequences of alternative observation and model choices, rather than selecting among them for a favorable diagnostic. Extending the calendar-screened sample into 2015 yields SIS and SVS correlations of 0.99 and 0.93 with the baseline. Applying return-magnitude cutoffs of 0.25, 0.35, or 0.50 changes some ranks but retains the same five leading baseline SIS institutions. Excluding Monte dei Paschi and Piraeus also preserves those five, with correlations of 0.95 and 0.93 across common banks. Omitting banks below EUR 1 billion in capitalization produces correlations of 0.99 for both scores.

Keeping all post-2015 aligned intervals produces lower correlations of 0.74 and 0.85. Using the full aligned unfiltered panel changes both the sample and the input scale and produces correlations of 0.52 and 0.75. This latter comparison illustrates why results from different processing pipelines should not be combined in one ranking or figure. It does not isolate a single source of change.

\begin{table}[!htbp]
\centering
\caption{Sample and specification sensitivities.}\label{tab:robustness}
\footnotesize
\begin{tabular}{p{7.5cm}rrrrr}
\toprule
Variant & T & d & Edges & $\rho$(SIS) & $\rho$(SVS) \\
\midrule
All aligned dates, unfiltered & 2204 & 52 & 196 & 0.52 & 0.75 \\
2016 onward, no calendar screen & 2180 & 52 & 199 & 0.74 & 0.85 \\
Calendar screen, including 2015 & 2012 & 52 & 215 & 0.99 & 0.93 \\
Return-magnitude screen: 0.25 & 1974 & 52 & 218 & 0.90 & 0.84 \\
Return-magnitude screen: 0.50 & 1988 & 52 & 219 & 0.94 & 0.89 \\
Return-magnitude screen: 0.35 & 1984 & 52 & 216 & 0.92 & 0.88 \\
Exclude BMPS and Piraeus & 1990 & 50 & 210 & 0.95 & 0.93 \\
VAR(1), standardized innovations & 1989 & 52 & 205 & 0.70 & 0.79 \\
Exclude capitalization below EUR 1bn & 1990 & 48 & 200 & 0.99 & 0.99 \\
\bottomrule
\end{tabular}
\notetext{Rank correlations compare common banks with the primary calendar-screened, volatility-standardized fit. Every row refits discovery and uses 1,000 coefficient resamples. The matched-observation VAR(0)/VAR(1) correlations are 0.71 and 0.79. The full aligned unfiltered row changes both filtering and observation selection; it is not a one-factor robustness comparison.}
\end{table}

For the VAR(1) sensitivity, the model is fitted on standardized observations with an intercept, and its innovations are rescaled before discovery. Relative to a VAR(0) fit on the same 1,989 dependent observations, SIS and SVS correlations are 0.71 and 0.79. The five leading source institutions are not identical: the VAR(1) ordering places CaixaBank, Intesa Sanpaolo, FinecoBank, Erste, and BBVA in the top five. Receiver top-five membership changes by one institution, with Banco BPM replacing Santander. These results support a relatively more stable receiving role while showing that source rankings depend on the treatment of short-run dynamics.

The influence exercises in Appendix Table~\ref{tab:influence} are complementary. Leaving out one quarter yields median SIS and SVS correlations of 0.85 and 0.88, with minima of 0.67 and 0.62. Removing one of the fifteen largest standardized observations yields medians of 0.91 and 0.94, with minima of 0.80 and 0.74. No conclusion about a specific institution should ignore these spreads.

The specification comparisons in Table~\ref{tab:robustness} assess the sensitivity of model-implied scores and rankings. Except where separate recovered-shock diagnostics are explicitly reported, these variants are treated as descriptive robustness exercises rather than as independently supported structural networks. Independence-support verdicts are specification-specific and are not inherited from the baseline or another fitted representation.

\section{Conclusion}
\label{sec:conclusion}
The analysis supports differentiated conclusions. SIS and SVS distinguish source and receiver sensitivities within the fitted network, and re-discovery of the ordering and coefficients, with the volatility filter held fixed, reveals greater aggregate reproducibility of receiver rankings. The temporal holdout additionally supports the relevance of selected pairs for subsequent-sample lower-tail association. Following the evidentiary distinctions of \citet[Section 3.3]{doannguyenbucciol2026}, these findings retain descriptive and exploratory value while their stronger interpretation remains conditional: the full-sample pairwise classifier rejects independence and the alternative-order comparison does not independently establish the recovered orientation. The scores are therefore reported as model-implied sensitivities, not as verified intervention effects, forecasts of future returns or measures of realized balance-sheet contagion.

The findings can be understood alongside theoretical models in which a bank's contribution to system-wide fragility depends on the distribution of links and the shocks arriving at particular locations in a network, not simply on its stand-alone size \citep{allen2000financial,acemoglu2015systemic,elliott2014financial}. The corresponding distinction between financial contagion and correlated market exposure is important here. An equity-price network can organize rapid common and directed movements that would be difficult to recover from bilateral regulatory exposures, while it cannot establish which transactions, creditor relationships or liquidity channels would transmit realized distress. SIS and SVS consequently provide complementary model-implied sensitivities, not substitute measures of the losses that might arise under a regulatory stress scenario.

The return construction and the available data impose additional limits. The calendar screen addresses visible multi-day intervals but cannot certify individual-stock trading continuity or vendor adjustments. Keeping large movements avoids treating magnitude itself as proof of error; issuer-exclusion and return-magnitude-screen sensitivities expose the consequences. The universe is a selected listed-bank panel, and capitalization weights are measured near the end of the sample. The paper consequently does not claim real-time supervisory rankings for the full European banking system.

Within this scope, the paper contributes a directed contemporaneous market-network representation and economically interpretable, distinct measures of systemic impact and vulnerability. By quantifying direct and indirect transmission and the reproducibility of source and receiver rankings, the framework adds a complementary perspective to conventional measures of connectedness and market-tail exposure. Independent shock measurements, richer exposure information, and higher-frequency data could strengthen identification, clarify the mechanisms underlying the estimated paths, and provide more discriminating tests of direction than the same-day residual-based tail contrast available here.

\FloatBarrier
\clearpage
\bibliography{Ref}

@article{RIGANA2023,
title = {A new way of measuring effects of financial crisis on contagion in currency markets},
journal = {International Review of Financial Analysis},
volume = {90},
pages = {102764},
year = {2023},
issn = {1057-5219},
doi = {https://doi.org/10.1016/j.irfa.2023.102764},
url = {https://www.sciencedirect.com/science/article/pii/S1057521923002806},
author = {Katerina Rigana and Ernst-Jan Camiel Wit and Samantha Cook},
}

@article{krause2012interbank,
  title={Interbank lending and the spread of bank failures: A network model of systemic risk},
  author={Krause, Andreas and Giansante, Simone},
  journal={Journal of Economic Behavior \& Organization},
  volume={83},
  number={3},
  pages={583--608},
  year={2012},
  publisher={Elsevier}
}

@article{castiglionesi2019financial,
  title={Financial integration and liquidity crises},
  author={Castiglionesi, Fabio and Feriozzi, Fabio and Lorenzoni, Guido},
  journal={Management Science},
  volume={65},
  number={3},
  pages={955--975},
  year={2019},
  publisher={INFORMS}
}

@article{castiglionesi2018liquidity,
  title={Liquidity flows in interbank networks},
  author={Castiglionesi, Fabio and Eboli, Mario},
  journal={Review of Finance},
  volume={22},
  number={4},
  pages={1291--1334},
  year={2018},
  publisher={Oxford University Press}
}

@article{glasserman2016contagion,
  title={Contagion in financial networks},
  author={Glasserman, Paul and Young, H Peyton},
  journal={Journal of Economic Literature},
  volume={54},
  number={3},
  pages={779--831},
  year={2016},
  publisher={American Economic Association 2014 Broadway, Suite 305, Nashville, TN 37203-2425}
}

@article{battiston2012debtrank,
  title={Debtrank: Too central to fail? financial networks, the fed and systemic risk},
  author={Battiston, Stefano and Puliga, Michelangelo and Kaushik, Rahul and Tasca, Paolo and Caldarelli, Guido},
  journal={Scientific reports},
  volume={2},
  number={1},
  pages={541},
  year={2012},
  publisher={Nature Publishing Group UK London}
}

@article{georg2013,
title = {The effect of the interbank network structure on contagion and common shocks},
journal = {Journal of Banking \& Finance},
volume = {37},
number = {7},
pages = {2216-2228},
year = {2013},
issn = {0378-4266},
doi = {https://doi.org/10.1016/j.jbankfin.2013.02.032},
url = {https://www.sciencedirect.com/science/article/pii/S0378426613001295},
author = {Co-Pierre Georg},
}

@article{battiston2012default,
  author  = {Battiston, Stefano and Delli Gatti, Domenico and Gallegati, Mauro and Greenwald, Bruce and Stiglitz, Joseph E.},
  title   = {Default Cascades: When Does Risk Diversification Increase Stability?},
  journal = {Journal of Financial Stability},
  year    = {2012},
  volume  = {8},
  number  = {3},
  pages   = {138--149},
  doi     = {10.1016/j.jfs.2012.01.002}
}

@article{acharya2025systemic,
  title={Systemic Risk Measures: From the Panic of 1907 to the Banking Stress of 2023},
  author={Acharya, Viral V and Brunnermeier, Markus K and Pierret, Diane},
  journal={Annual Review of Financial Economics},
  volume={17},
  year={2025},
  publisher={Annual Reviews}
}

@article{elliott2014financial,
  title={Financial networks and contagion},
  author={Elliott, Matthew and Golub, Benjamin and Jackson, Matthew O},
  journal={American Economic Review},
  volume={104},
  number={10},
  pages={3115--3153},
  year={2014},
  publisher={American Economic Association 2014 Broadway, Suite 305, Nashville, TN 37203}
}

@article{shu2025endogenous,
  title={Endogenous Risk Exposure and Systemic Instability},
  author={Shu, Chong},
  journal={Management Science},
  volume={71},
  number={7},
  pages={5511--5528},
  year={2025},
  publisher={INFORMS}
}

@article{acemoglu2015systemic,
  title={Systemic risk and stability in financial networks},
  author={Acemoglu, Daron and Ozdaglar, Asuman and Tahbaz-Salehi, Alireza},
  journal={American Economic Review},
  volume={105},
  number={2},
  pages={564--608},
  year={2015},
  publisher={American Economic Association 2014 Broadway, Suite 305, Nashville, TN 37203}
}

@article{siebenbrunner2021quantifying,
  title={Quantifying the importance of different contagion channels as sources of systemic risk},
  author={Siebenbrunner, Christoph},
  journal={Journal of Economic Interaction and Coordination},
  volume={16},
  number={1},
  pages={103--131},
  year={2021},
  publisher={Springer}
}

@article{sui2020financial,
  title={Financial contagion in a core-periphery interbank network},
  author={Sui, Peng and Tanna, Sailesh and Zhou, Dandan},
  journal={The European Journal of Finance},
  volume={26},
  number={7-8},
  pages={691--710},
  year={2020},
  publisher={Taylor \& Francis}
}

@article{cabrales2017risk,
  title={Risk sharing and contagion in networks},
  author={Cabrales, Antonio and Gottardi, Piero and Vega-Redondo, Fernando},
  journal={The Review of Financial Studies},
  volume={30},
  number={9},
  pages={3086--3127},
  year={2017},
  publisher={Oxford University Press}
}

@article{brownlees2017srisk,
  title={SRISK: A conditional capital shortfall measure of systemic risk},
  author={Brownlees, Christian and Engle, Robert F},
  journal={The Review of Financial Studies},
  volume={30},
  number={1},
  pages={48--79},
  year={2017},
  publisher={Oxford University Press}
}

@article{Moneta_2013,
title = {Causal Inference by Independent Component Analysis: Theory and Applications},
journal = {Oxford Bulletin of Economics and Statistics},
year = {2013},
doi = {https://doi.org/10.1111/j.1468-0084.2012.00710.x},
url = {https://onlinelibrary.wiley.com/doi/abs/10.1111/j.1468-0084.2012.00710.x},
author = {Alessio Moneta and Doris Entner and Patrik O Hoyer and Alex Coad},
volume = {75},
number = {5},
pages = {705--730}
}

@article{benoit2017risks,
  title={Where the risks lie: A survey on systemic risk},
  author={Benoit, Sylvain and Colliard, Jean-Edouard and Hurlin, Christophe and P{\'e}rignon, Christophe},
  journal={Review of Finance},
  volume={21},
  number={1},
  pages={109--152},
  year={2017},
  publisher={Oxford University Press}
}

@article{jackson2025optimal,
  title={Optimal regulation and investment incentives in financial networks},
  author={Jackson, Matthew O and Pernoud, Agathe},
  journal={arXiv preprint arXiv:2506.16648},
  year={2025}
}

@article{billio2012econometric,
  title={Econometric measures of connectedness and systemic risk in the finance and insurance sectors},
  author={Billio, Monica and Getmansky, Mila and Lo, Andrew W and Pelizzon, Loriana},
  journal={Journal of financial economics},
  volume={104},
  number={3},
  pages={535--559},
  year={2012},
  publisher={Elsevier}
}

@article{silva2017analysis,
  title={An analysis of the literature on systemic financial risk: A survey},
  author={Silva, Walmir and Kimura, Herbert and Sobreiro, Vinicius Amorim},
  journal={Journal of Financial Stability},
  volume={28},
  pages={91--114},
  year={2017},
  publisher={Elsevier}
}

@techreport{montagna2020origin,
  author      = {Mattia Montagna and Gabriele Torri and Giovanni Covi},
  title       = {On the Origin of Systemic Risk},
  institution = {European Central Bank},
  series      = {ECB Working Paper Series},
  number      = {2502},
  year        = {2020},
  url         = {https://www.ecb.europa.eu/pub/pdf/scpwps/ecb.wp2502~b7a6f6d2c7.en.pdf}
}

@article{GAO2018,
title = {Causal data science for financial stress testing},
journal = {Journal of Computational Science},
volume = {26},
pages = {294-304},
year = {2018},
issn = {1877-7503},
doi = {https://doi.org/10.1016/j.jocs.2018.04.003},
url = {https://www.sciencedirect.com/science/article/pii/S1877750317311377},
author = {Gelin Gao and Bud Mishra and Daniele Ramazzotti},
}

@misc{kumar2023,
      title={Causal Inference for Banking Finance and Insurance A Survey}, 
      author={Satyam Kumar and Yelleti Vivek and Vadlamani Ravi and Indranil Bose},
      year={2023},
      eprint={2307.16427},
      archivePrefix={arXiv},
      primaryClass={cs.AI},
      url={https://arxiv.org/abs/2307.16427} 
}

@book{Peters2017,
author = {Peters, J. and Janzing, D. and Sch\"olkopf, B.},
title = {Elements of Causal Inference: Foundations and Learning Algorithms},
address = {Cambridge, MA, USA},
publisher = {MIT Press},
year = {2017}
}

@incollection{billio2017systemic,
  title={Systemic risk and financial interconnectedness: network measures and the impact of the indirect effect},
  author={Billio, Monica and Costola, Michele and Panzica, Roberto and Pelizzon, Loriana},
  booktitle={Systemic Risk Tomography},
  pages={43--72},
  year={2017},
  publisher={Elsevier}
}

@article{hyvarinen2010,
  author  = {Aapo Hyv{{\"a}}rinen and Kun Zhang and Shohei Shimizu and Patrik O. Hoyer},
  title   = {Estimation of a Structural Vector Autoregression Model Using Non-{G}aussianity},
  journal = {Journal of Machine Learning Research},
  year    = {2010},
  volume  = {11},
  number  = {56},
  pages   = {1709--1731},
  url     = {http://jmlr.org/papers/v11/hyvarinen10a.html}
}

@article{kanno2015assessing,
  title={Assessing systemic risk using interbank exposures in the global banking system},
  author={Kanno, Masayasu},
  journal={Journal of Financial Stability},
  volume={20},
  pages={105--130},
  year={2015},
  publisher={Elsevier}
}

@misc{jiang2023,
      title={Linkages among the Foreign Exchange, Stock, and Bond Markets in Japan and the United States}, 
      author={Yi Jiang and Shohei Shimizu},
      year={2023},
      eprint={2310.16841},
      archivePrefix={arXiv},
      primaryClass={q-fin.ST},
      url={https://arxiv.org/abs/2310.16841}, 
}

@inproceedings{xu2014,
  author       = {Xu, Xiaojie},
  title        = {Causality and Price Discovery in U.S. Corn Markets: An Application of Error Correction Modeling and Directed Acyclic Graphs},
  booktitle    = {2014 Annual Meeting, July 27-29, 2014, Minneapolis, Minnesota},
  year     = {2014},
  journal = {Agricultural and Applied Economics Association},
  pages        = {1--85},
  doi          = {10.22004/ag.econ.169806},
  url          = {https://ageconsearch.umn.edu/record/169806}
}

@article{Shimizu06a,
  author  = {Shohei Shimizu and Patrik O. Hoyer and Aapo Hyvrinen and Antti Kerminen},
  title   = {A Linear Non-{G}aussian Acyclic Model for Causal Discovery},
  journal = {Journal of Machine Learning Research},
  year    = {2006},
  volume  = {7},
  number  = {72},
  pages   = {2003--2030},
  url     = {http://jmlr.org/papers/v7/shimizu06a.html}
}

@article{ciarli2023does,
  title={Does exporting cause productivity growth? Evidence from Chilean firms},
  author={Ciarli, Tommaso and Coad, Alex and Moneta, Alessio},
  journal={Structural Change and Economic Dynamics},
  volume={66},
  pages={228--239},
  year={2023},
  publisher={Elsevier}
}

@article{halaj2025decomposing,
  title={Decomposing systemic risk: The roles of contagion and common exposures},
  author={Ha{\l}aj, Grzegorz and Hipp, Ruben},
  journal={Journal of Financial Stability},
  pages={101451},
  year={2025},
  publisher={Elsevier}
}

@article{allen2000financial,
  title={Financial contagion},
  author={Allen, Franklin and Gale, Douglas},
  journal={Journal of political economy},
  volume={108},
  number={1},
  pages={1--33},
  year={2000},
  publisher={The University of Chicago Press}
}

@article{GAO2017,
title = {Efficient Simulation of Financial Stress Testing Scenarios with {Suppes-Bayes} Causal Networks},
journal = {Procedia Computer Science},
volume = {108},
pages = {272-284},
year = {2017},
note = {International Conference on Computational Science, ICCS 2017, 12-14 June 2017, Zurich, Switzerland},
issn = {1877-0509},
doi = {https://doi.org/10.1016/j.procs.2017.05.167},
url = {https://www.sciencedirect.com/science/article/pii/S1877050917307500},
author = {Gelin Gao and Bud Mishra and Daniele Ramazzotti}
}

@article{moneta2004identification,
  author  = {Moneta, Alessio},
  title   = {Identification of Monetary Policy Shocks: A Graphical Causal Approach},
  journal = {Notas Económicas},
  volume  = {20},
  pages   = {39--62},
  year    = {2004},
  publisher = {Faculty of Economics, University of Coimbra},
  doi     = {10.14195/2183-203X_20_3},
  url     = {https://doi.org/10.14195/2183-203X_20_3}
}

@book{pearl2000causality,
  title     = {Causality: Models, Reasoning, and Inference},
  author    = {Pearl, Judea},
  year      = {2000},
  publisher = {Cambridge University Press},
  address   = {Cambridge, UK},
  isbn      = {9780521773621},
  url       = {https://doi.org/10.1017/CBO9780511803161}
}

@book{pearl2009causality,
  author    = {Pearl, Judea},
  title     = {Causality: Models, Reasoning, and Inference},
  edition   = {2},
  year      = {2009},
  publisher = {Cambridge University Press},
  address   = {Cambridge, UK},
  isbn      = {9780521895606},
  url       = {https://doi.org/10.1017/CBO9780511803161}
}

@article{glymour2019review,
  author    = {Clark Glymour and Kun Zhang and Peter Spirtes},
  title     = {Review of Causal Discovery Methods Based on Graphical Models},
  journal   = {Frontiers in Genetics},
  volume    = {10},
  pages     = {524},
  year      = {2019},
  doi       = {10.3389/fgene.2019.00524},
  url       = {https://doi.org/10.3389/fgene.2019.00524}
}

@article{comon1994independent,
  title={Independent component analysis, a new concept?},
  author={Comon, Pierre},
  journal={Signal processing},
  volume={36},
  number={3},
  pages={287--314},
  year={1994},
  publisher={Elsevier}
}

@article{hyvarinen2013independent,
  title={Independent component analysis: recent advances},
  author={Hyv{\"a}rinen, Aapo},
  journal={Philosophical Transactions of the Royal Society A: Mathematical, Physical and Engineering Sciences},
  volume={371},
  number={1984},
  pages={20110534},
  year={2013},
  publisher={The Royal Society Publishing}
}

@inproceedings{Gretton_2005_HSIC,
  author    = {Arthur Gretton and Olivier Bousquet and Alex Smola and Bernhard Sch{\"o}lkopf},
  title     = {Measuring Statistical Dependence with Hilbert-Schmidt Norms},
  booktitle = {Algorithmic Learning Theory},
  series    = {Lecture Notes in Computer Science},
  volume    = {3734},
  pages     = {63--77},
  year      = {2005},
  publisher = {Springer},
  doi       = {10.1007/11564089_7},
  url       = {https://doi.org/10.1007/11564089_7}
}

@inproceedings{Gretton_2007_HSIC_Test,
  author    = {Arthur Gretton and Kenji Fukumizu and Choon Hui Teo and Le Song and Bernhard Sch{\"o}lkopf and Alex J. Smola},
  title     = {A Kernel Statistical Test of Independence},
  booktitle = {Advances in Neural Information Processing Systems 20},
  pages     = {585--592},
  year      = {2007},
  url       = {https://proceedings.neurips.cc/paper_files/paper/2007/hash/d5cfead94f5350c12c322b5b664544c1-Abstract.html}
}

@article{diebold2014network,
  author  = {Francis X. Diebold and Kamil Yilmaz},
  title   = {On the Network Topology of Variance Decompositions: Measuring the Connectedness of Financial Firms},
  journal = {Journal of Econometrics},
  year    = {2014},
  volume  = {182},
  number  = {1},
  pages   = {119--134},
  doi     = {10.1016/j.jeconom.2014.04.012},
  url     = {https://doi.org/10.1016/j.jeconom.2014.04.012}
}

@article{pesaran1998generalized,
  author  = {M. Hashem Pesaran and Yongcheol Shin},
  title   = {Generalized Impulse Response Analysis in Linear Multivariate Models},
  journal = {Economics Letters},
  year    = {1998},
  volume  = {58},
  number  = {1},
  pages   = {17--29},
  doi     = {10.1016/S0165-1765(97)00214-0},
  url     = {https://doi.org/10.1016/S0165-1765(97)00214-0}
}

@article{shimizu2011directlingam,
  author  = {Shohei Shimizu and Takanori Inazumi and Yasuhiro Sogawa and Aapo Hyv{\"a}rinen and Yoshinobu Kawahara and Takashi Washio and Patrik O. Hoyer and Kenneth Bollen},
  title   = {{DirectLiNGAM}: A Direct Method for Learning a Linear Non-{G}aussian Structural Equation Model},
  journal = {Journal of Machine Learning Research},
  year    = {2011},
  volume  = {12},
  number  = {33},
  pages   = {1225--1248},
  url     = {https://jmlr.org/papers/v12/shimizu11a.html}
}

@article{hyvarinen2013pairwise,
  author  = {Aapo Hyv{\"a}rinen and Stephen M. Smith},
  title   = {Pairwise Likelihood Ratios for Estimation of Non-{G}aussian Structural Equation Models},
  journal = {Journal of Machine Learning Research},
  year    = {2013},
  volume  = {14},
  number  = {4},
  pages   = {111--152},
  url     = {https://jmlr.org/papers/v14/hyvarinen13a.html}
}

@article{bae2003coexceedance,
  author  = {Bae, Kee-Hong and Karolyi, G. Andrew and Stulz, Ren{\'e} M.},
  title   = {A New Approach to Measuring Financial Contagion},
  journal = {The Review of Financial Studies},
  year    = {2003},
  volume  = {16},
  number  = {3},
  pages   = {717--763},
  doi     = {10.1093/rfs/hhg012}
}

@article{baur2005coexceedances,
  author  = {Baur, Dirk G. and Schulze, Niels},
  title   = {Coexceedances in Financial Markets---A Quantile Regression Analysis of Contagion},
  journal = {Emerging Markets Review},
  year    = {2005},
  volume  = {6},
  number  = {1},
  pages   = {21--43},
  doi     = {10.1016/j.ememar.2004.10.001}
}

@article{chiu2015industry,
  author  = {Chiu, Wan-Chien and Pe{\~n}a, Juan Ignacio and Wang, Chih-Wei},
  title   = {Industry Characteristics and Financial Risk Contagion},
  journal = {Journal of Banking \& Finance},
  year    = {2015},
  volume  = {50},
  pages   = {411--427},
  doi     = {10.1016/j.jbankfin.2014.04.003}
}

@article{acharya2017, author={Acharya, Viral V. and Pedersen, Lasse H. and Philippon, Thomas and Richardson, Matthew}, title={Measuring systemic risk}, journal={Review of Financial Studies}, volume={30}, number={1}, pages={2--47}, year={2017}}

@article{chordia2000, author={Chordia, Tarun and Swaminathan, Bhaskaran}, title={Trading volume and cross-autocorrelations in stock returns}, journal={Journal of Finance}, volume={55}, number={2}, pages={913--935}, year={2000}}

@article{hasbrouck1995, author={Hasbrouck, Joel}, title={One security, many markets: Determining the contributions to price discovery}, journal={Journal of Finance}, volume={50}, number={4}, pages={1175--1199}, year={1995}}

@article{burns1998, author={Burns, Patrick and Engle, Robert F. and Mezrich, Joseph}, title={Correlations and volatilities of asynchronous data}, journal={Journal of Derivatives}, volume={5}, number={4}, pages={7--18}, year={1998}}

@incollection{komatsu2010, author={Komatsu, Yusuke and Shimizu, Shohei and Shimodaira, Hidetoshi}, title={Assessing statistical reliability of {LiNGAM} via multiscale bootstrap}, booktitle={Artificial Neural Networks -- {ICANN} 2010}, series={Lecture Notes in Computer Science}, volume={6354}, pages={309--314}, publisher={Springer}, address={Berlin, Heidelberg}, year={2010}}

@article{martens2001, author={Martens, Martin and Poon, Ser-Huang}, title={Returns synchronization and daily correlation dynamics between international stock markets}, journal={Journal of Banking \& Finance}, volume={25}, number={10}, pages={1805--1827}, year={2001}}

@article{lo1990, author={Lo, Andrew W. and MacKinlay, A. Craig}, title={When are contrarian profits due to stock market overreaction?}, journal={Review of Financial Studies}, volume={3}, number={2}, pages={175--205}, year={1990}}

@article{montielolea2022, author={Montiel Olea, Jos\'e Luis and Plagborg-M{\o}ller, Mikkel and Qian, Eric}, title={{SVAR} identification from higher moments: Has the simultaneous causality problem been solved?}, journal={AEA Papers and Proceedings}, volume={112}, pages={481--485}, year={2022}}

@article{hautsch2015financial,
  author  = {Nikolaus Hautsch and Julia Schaumburg and Melanie Schienle},
  title   = {Financial Network Systemic Risk Contributions},
  journal = {Review of Finance},
  year    = {2015},
  volume  = {19},
  number  = {2},
  pages   = {685--738},
  doi     = {10.1093/rof/rfu010},
  url     = {https://doi.org/10.1093/rof/rfu010}
}

@article{hautsch2014forecasting,
  author  = {Nikolaus Hautsch and Julia Schaumburg and Melanie Schienle},
  title   = {Forecasting Systemic Impact in Financial Networks},
  journal = {International Journal of Forecasting},
  year    = {2014},
  volume  = {30},
  number  = {3},
  pages   = {781--794},
  doi     = {10.1016/j.ijforecast.2013.09.004},
  url     = {https://doi.org/10.1016/j.ijforecast.2013.09.004}
}

@article{betz2016systemic,
  author  = {Frank Betz and Nikolaus Hautsch and Tuomas A. Peltonen and Melanie Schienle},
  title   = {Systemic Risk Spillovers in the {European} Banking and Sovereign Network},
  journal = {Journal of Financial Stability},
  year    = {2016},
  volume  = {25},
  pages   = {206--224},
  doi     = {10.1016/j.jfs.2015.10.006},
  url     = {https://doi.org/10.1016/j.jfs.2015.10.006}
}

@incollection{ilmonen2010new,
  author    = {Pauliina Ilmonen and Klaus Nordhausen and Hannu Oja and Esa Ollila},
  title     = {A New Performance Index for {ICA}: Properties, Computation and Asymptotic Analysis},
  booktitle = {Latent Variable Analysis and Signal Separation},
  editor    = {V. Vigneron and V. Zarzoso and E. Moreau and R. Gribonval and E. Vincent},
  series    = {Lecture Notes in Computer Science},
  volume    = {6365},
  pages     = {229--236},
  publisher = {Springer},
  year      = {2010},
  doi       = {10.1007/978-3-642-15995-4_29},
  url       = {https://doi.org/10.1007/978-3-642-15995-4_29}
}

@misc{doannguyenbucciol2026,
  author       = {Doan Nguyen, Tu Anh and Bucciol, Alessandro},
  title        = {Persistence is not independence: A structural credibility protocol for distribution-based recovery in {VAR}--{LiNGAM}},
  year         = {2026},
  howpublished = {arXiv preprint arXiv:2610.25238},
  doi          = {10.48550/arXiv.2610.25238}
}
\clearpage
\appendix

\section{Sample institutions}
\label{app:data}
Table~\ref{tab:banks} records the analysis universe and the capitalization weights. These are fixed weights, not reconstructed point-in-time market capitalizations. The sample-selection and return-magnitude-screen sensitivities are documented in Section~\ref{sec:var1_robustness}; the observation-level audit is retained in the replication materials, rather than reproduced here.

\small
\begin{longtable}{rlp{5.1cm}lr}
\caption{Sample banks and market capitalizations (EUR billion, 1 July 2025).}\label{tab:banks}\\
\toprule No. & Ticker & Institution & Country & Cap. \\ \midrule\endfirsthead
\multicolumn{5}{l}{\small Table \thetable\ continued}\\ \toprule No. & Ticker & Institution & Country & Cap. \\ \midrule\endhead
\bottomrule\endfoot
1 & A5G.IR & AIB Group plc & Ireland & 14.509 \\
2 & ABN.AS & ABN AMRO Bank N.V. & Netherlands & 19.377 \\
3 & ACA.PA & Crédit Agricole S.A. & France & 48.248 \\
4 & ALPHA.AT & Alpha Bank S.A. & Greece & 6.908 \\
5 & BAMI.MI & Banco BPM S.p.A. & Italy & 14.774 \\
6 & BBVA.MC & Banco Bilbao Vizcaya Argentaria & Spain & 73.973 \\
7 & BCP.LS & Banco Comercial Português & Portugal & 9.814 \\
8 & BGN.MI & Banca Generali S.p.A. & Italy & 5.294 \\
9 & BIRG.IR & Bank of Ireland Group plc & Ireland & 11.478 \\
10 & BKT.MC & Bankinter, S.A. & Spain & 9.930 \\
11 & BMED.MI & Banca Mediolanum S.p.A. & Italy & 10.554 \\
12 & BMPS.MI & Monte dei Paschi di Siena & Italy & 8.908 \\
13 & BNP.PA & BNP Paribas S.A. & France & 84.983 \\
14 & BPE.MI & BPER Banca S.p.A. & Italy & 10.586 \\
15 & BPSO.MI & Banca Popolare di Sondrio & Italy & 5.143 \\
16 & BST.MI & Banca Sistema S.p.A. & Italy & 0.148 \\
17 & BWB.DE & Baader Bank AG & Germany & 0.212 \\
18 & CABK.MC & CaixaBank, S.A. & Spain & 51.917 \\
19 & CBK.DE & Commerzbank AG & Germany & 30.104 \\
20 & CE.MI & Credito Emiliano & Italy & 4.197 \\
21 & DANSKE.CO & Danske Bank A/S & Denmark & 28.245 \\
22 & DBK.DE & Deutsche Bank AG & Germany & 47.083 \\
23 & EBS.VI & Erste Group Bank AG & Austria & 27.809 \\
24 & ETE.AT & National Bank of Greece & Greece & 10.025 \\
25 & EUROB.AT & Eurobank Ergasias & Greece & 10.839 \\
26 & FBK.MI & FinecoBank S.p.A. & Italy & 11.357 \\
27 & GLE.PA & Société Générale & France & 37.231 \\
28 & IF.MI & Banca IFIS & Italy & 1.183 \\
29 & ING.WA & ING Bank Śląski & Poland & 9.486 \\
30 & INGA.AS & ING Groep N.V. & Netherlands & 56.062 \\
31 & ISP.MI & Intesa Sanpaolo & Italy & 85.512 \\
32 & JYSK.CO & Jyske Bank A/S & Denmark & 5.168 \\
33 & KBC.BR & KBC Group N.V. & Belgium & 34.398 \\
34 & KOMB.PR & Komerční banka & Czech Republic & 7.788 \\
35 & MB.MI & Mediobanca & Italy & 15.543 \\
36 & MBK.WA & mBank S.A. & Poland & 7.932 \\
37 & NDA-FI.HE & Nordea Bank Abp & Finland & 41.191 \\
38 & OTP.BD & OTP Bank Nyrt & Hungary & 17.414 \\
39 & PBB.DE & Pfandbriefbank AG & Germany & 0.707 \\
40 & PEO.WA & Bank Pekao S.A. & Poland & 11.269 \\
41 & PKO.WA & PKO Bank Polski & Poland & 20.000 \\
42 & PRO.MI & Banca Profilo S.p.A. & Italy & 0.110 \\
43 & RBI.VI & Raiffeisen Intl. & Austria & 8.315 \\
44 & SAB.MC & Banco Sabadell & Spain & 13.973 \\
45 & SAN.MC & Banco Santander & Spain & 103.543 \\
46 & SEB-A.ST & SEB AB & Sweden & 29.896 \\
47 & SHB-A.ST & Handelsbanken & Sweden & 22.450 \\
48 & SPL.WA & Santander Polska & Poland & 11.707 \\
49 & SYDB.CO & Sydbank A/S & Denmark & 3.129 \\
50 & TPEIR.AT & Piraeus Bank & Greece & 4.447 \\
51 & UCG.MI & UniCredit S.p.A. & Italy & 86.580 \\
52 & EXSA.DE & STOXX Europe 600 tracker & Market & --- \\
\end{longtable}\normalsize
\noindent{\footnotesize\textit{Ticker-label convention.} For readability, plot labels use periods in place of the hyphens in three input tickers: NDA.FI.HE, SEB.A.ST and SHB.A.ST in the figures correspond, respectively, to NDA-FI.HE, SEB-A.ST and SHB-A.ST in the bank table.\par}

\FloatBarrier

\section{Additional discovery and influence diagnostics}
\label{app:discovery}
Figure~\ref{fig:frequency} displays direction frequencies for all ordered pairs with banks in alphabetical order and the market proxy last. It includes directions absent from the baseline graph but selected in some bootstrap samples. Table~\ref{tab:backbone_full} lists all \Nbackbone{} edges of the strict backbone; Table~\ref{tab:backbone} in the main text shows the sixteen largest. Table~\ref{tab:influence} summarizes leave-one-quarter and leave-one-extreme-observation fits. Detailed fitted ranks and omitted dates are included in the computational outputs.

\begin{table}[!htbp]
\centering
\caption{All edges of the strict discovery backbone.}\label{tab:backbone_full}
\small
\begin{tabular}{rllrrr}
\toprule
No. & Source & Target & Coefficient & Frequency & Coef.\ $\times$ freq. \\
\midrule
1 & EXSA.DE & FBK.MI & 0.550 & 0.93 & 0.513 \\
2 & JYSK.CO & SYDB.CO & 0.446 & 0.96 & 0.427 \\
3 & BNP.PA & ACA.PA & 0.421 & 0.96 & 0.405 \\
4 & INGA.AS & ABN.AS & 0.382 & 0.99 & 0.379 \\
5 & ETE.AT & ALPHA.AT & 0.428 & 0.84 & 0.359 \\
6 & BIRG.IR & A5G.IR & 0.350 & 0.93 & 0.326 \\
7 & CABK.MC & SAB.MC & 0.360 & 0.89 & 0.322 \\
8 & EUROB.AT & ALPHA.AT & 0.377 & 0.85 & 0.319 \\
9 & FBK.MI & BGN.MI & 0.315 & 0.85 & 0.269 \\
10 & BNP.PA & GLE.PA & 0.300 & 0.90 & 0.269 \\
11 & ISP.MI & UCG.MI & 0.302 & 0.86 & 0.260 \\
12 & CABK.MC & BBVA.MC & 0.302 & 0.82 & 0.248 \\
13 & ETE.AT & TPEIR.AT & 0.287 & 0.83 & 0.238 \\
14 & ISP.MI & BAMI.MI & 0.256 & 0.86 & 0.221 \\
15 & EXSA.DE & BGN.MI & 0.229 & 0.95 & 0.218 \\
16 & EUROB.AT & TPEIR.AT & 0.225 & 0.86 & 0.193 \\
17 & BKT.MC & SAB.MC & 0.196 & 0.89 & 0.175 \\
18 & DANSKE.CO & SYDB.CO & 0.191 & 0.86 & 0.164 \\
19 & EXSA.DE & SEB-A.ST & 0.170 & 0.95 & 0.161 \\
20 & EXSA.DE & BBVA.MC & 0.177 & 0.89 & 0.158 \\
21 & EXSA.DE & OTP.BD & 0.180 & 0.87 & 0.157 \\
22 & EXSA.DE & NDA-FI.HE & 0.149 & 1.00 & 0.149 \\
23 & ISP.MI & BPSO.MI & 0.143 & 0.88 & 0.126 \\
24 & EXSA.DE & BWB.DE & 0.146 & 0.85 & 0.124 \\
25 & EXSA.DE & SAN.MC & 0.133 & 0.93 & 0.124 \\
26 & EXSA.DE & BAMI.MI & -0.143 & 0.85 & -0.122 \\
27 & EXSA.DE & PBB.DE & 0.140 & 0.84 & 0.117 \\
28 & BKT.MC & KBC.BR & 0.128 & 0.80 & 0.103 \\
29 & EXSA.DE & PKO.WA & 0.112 & 0.80 & 0.090 \\
30 & EXSA.DE & BMED.MI & 0.102 & 0.83 & 0.085 \\
31 & KOMB.PR & EBS.VI & 0.078 & 0.84 & 0.066 \\
32 & EXSA.DE & BNP.PA & 0.072 & 0.90 & 0.065 \\
\bottomrule
\end{tabular}
\notetext{The strict backbone contains the 32 baseline edges whose direction is recovered in at least 80\% of 500 re-discovery replications. Coefficient is the selected full-sample coefficient; frequency is the share of replications in which the same directed edge is selected. Rows are ordered by the absolute product of the two, as in Table~\ref{tab:backbone}. Tickers are defined in Table~\ref{tab:banks}.}
\end{table}

\figone{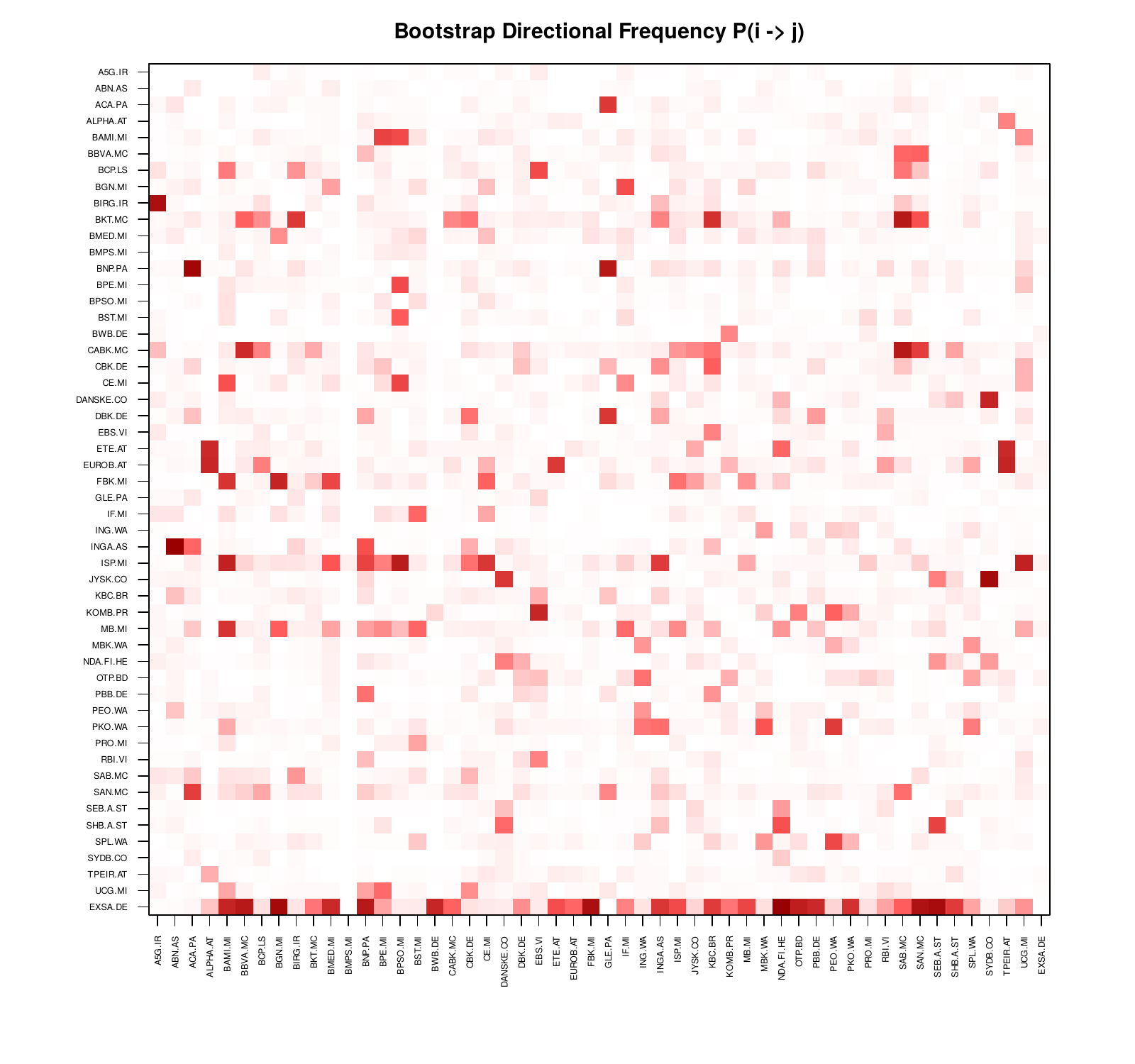}{Direction frequencies across all re-discovered graphs. The matrix uses the alphabetical ordering of banks, with the market proxy last; source rows and target columns are shown. Frequencies above the diagonal can occur because the ordering is re-estimated rather than fixed.}{fig:frequency}{0.94}

\begin{table}[!htbp]
\centering
\caption{Influence of quarters and extreme standardized observations.}\label{tab:influence}
\small
\begin{tabular}{lrrrrr}
\toprule
Deletion scheme & Fits & SIS med. & SIS min. & SVS med. & SVS min. \\
\midrule
Leave one quarter out & 38 & 0.85 & 0.67 & 0.88 & 0.62 \\
Leave one extreme session out & 15 & 0.91 & 0.80 & 0.94 & 0.74 \\
\bottomrule
\end{tabular}
\notetext{Each omission refits the GARCH filter and discovery step, with 500 inner coefficient resamples. Correlations are relative to the primary fit. Extreme sessions are the fifteen largest cross-sectional absolute standardized returns; this diagnostic does not delete them from the baseline.}
\end{table}

\FloatBarrier

\section{Why a held-out directional gap does not identify direction}
\label{app:counterexample}
\figone{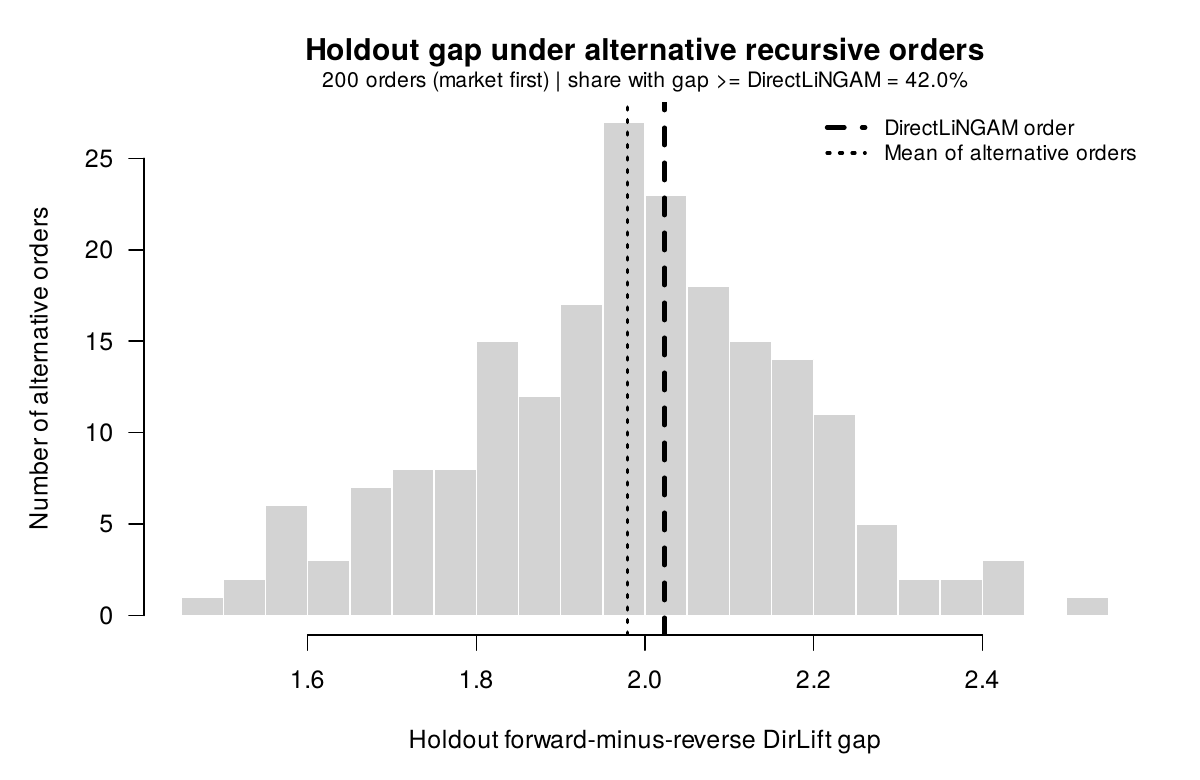}{Holdout forward-minus-reverse tail-lift gaps under 200 alternative recursive orders with refitted coefficients, networks and shock transformations. The DirectLiNGAM gap is not exceptional within this comparator; this is not an exact causal-null randomization test.}{fig:coherent}{0.85}

Consider an unobserved common factor and two independent idiosyncratic disturbances:
\begin{equation}
 X_1=\sqrt{\rho}\,F+\sqrt{1-\rho}\,U,\qquad
 X_2=\sqrt{\rho}\,F+\sqrt{1-\rho}\,V,
 \qquad 0<\rho<1,
\end{equation}
where $F,U,V$ are independent standard Gaussian variables. There is no structural arrow from $X_1$ to $X_2$ or from $X_2$ to $X_1$; their correlation is generated by the common cause.

A recursive regression ordered as $1\rightarrow2$ nevertheless writes
\begin{equation}
 \widetilde\varepsilon_1=X_1,\qquad
 \widetilde\varepsilon_2=X_2-\rho X_1.
\end{equation}
Joint Gaussianity makes $\widetilde\varepsilon_2$ independent of $X_1$. A lower-tail event in $\widetilde\varepsilon_1=X_1$ raises the probability of a lower-tail event in $X_2$, while a lower-tail event in $\widetilde\varepsilon_2$ does not change the lower-tail probability of $X_1$. Thus the forward lift exceeds one and the reverse lift equals one. The same argument applies after reversing the recursive order. Since the population is stable, both gaps persist when the regression is fitted on training data and evaluated on a fresh independent sample.

The accompanying numerical illustration uses $\rho=0.6$, 100,000 training observations and 100,000 held-out observations. The two orders produce forward lifts of approximately 6.21 and reverse lifts near one. This Gaussian, confounded example lies outside the independent non-Gaussian causally sufficient model required for LiNGAM identification. It does not refute that theorem. It shows that a positive directional-lift gap, even in held-out data, cannot by itself verify those assumptions or identify the arrow. The coherently re-estimated empirical comparison in Section~\ref{sec:tail_results} addresses the same diagnostic limitation within the banking data.

\section{Simulation design and finite-sample filtering comparison}
\label{app:mc}
The simulation uses the selected 52-variable acyclic matrix as a fixed coefficient design. For each of 50 replications per design, we generate 1,990 retained observations after a burn-in of 500. The disturbances have variance-standardized Student-$t_5$ innovations. The three designs are: independent identically distributed shocks; component-specific GARCH(1,1) shocks; and the latter shocks multiplied by a common stochastic volatility component. For the GARCH designs, coefficients vary linearly across components from $\alpha=0.04$ to $0.12$ and from $\beta=0.88$ to $0.84$, with $\omega=1-\alpha-\beta$. The common log-volatility component follows an AR(1) with coefficient 0.95 and innovation standard deviation 0.15.

Within every replication, DirectLiNGAM is applied both to standardized unfiltered observations and to observations processed by the fitted univariate GARCH filters. Directed links are selected using the 0.05 coefficient threshold, without a nested Wald bootstrap. Precision is the fraction of selected directed edges present in the generating graph; recall is the fraction of generating edges recovered in the correct direction. Table~\ref{tab:mc} reports means across replications. These are edge-recovery diagnostics for the specified simulation designs, not an empirical test of the true banking network.

The common-volatility design shows an improvement in mean precision after filtering. The other designs show similar performance before and after filtering. This finite-sample evidence does not prove that univariate filtering preserves a constant structural coefficient matrix under arbitrary heterogeneous volatility, as equation~\eqref{eq:scale} makes clear. It does not account for unobserved confounding or establish the identifying assumptions in the empirical sample.

\begin{table}[!htbp]
\centering
\caption{Directed-edge recovery in controlled simulation designs.}\label{tab:mc}
\small
\begin{tabular}{llrr}
\toprule
Design & Input & Precision & Recall \\
\midrule
Independent Student-$t$ & Unfiltered & 0.768 & 0.881 \\
Independent Student-$t$ & Filtered & 0.770 & 0.883 \\
Independent GARCH & Unfiltered & 0.775 & 0.886 \\
Independent GARCH & Filtered & 0.773 & 0.886 \\
Common volatility & Unfiltered & 0.679 & 0.857 \\
Common volatility & Filtered & 0.721 & 0.863 \\
\bottomrule
\end{tabular}
\notetext{Means across 50 replications per design. Each replication has 1,990 observations and 52 variables. Selection here uses only the 0.05 coefficient threshold, not the empirical nested Wald procedure. These experiments do not establish that the maintained empirical model is correctly specified.}
\end{table}

\section{Recursiveness diagnostics}
\label{app:recursiveness}
\paragraph{ICA--Cholesky comparison.} FastICA uses the log-cosh contrast with symmetric decorrelation, tolerance $10^{-5}$ and at most 4,000 iterations. Its components have unit sample variance, so $\widetilde{\bm{C}}\widetilde{\bm{C}}'=\bm{\Sigma}^{*}$ holds to numerical precision when $\bm{\Sigma}^{*}$ uses the divisor $n$. With the divisor $n-1$, the aligned distance changes in the third decimal and the comparisons are unchanged. The minimum in equation~\eqref{eq:recursive_distance} assigns the columns of $\widetilde{\bm{C}}$ to those of $\bm{D}$ so as to maximize the sum of absolute inner products, a linear assignment problem, and sets each sign to that of the corresponding inner product. The minimum-distance index of \citet{ilmonen2010new} is $(p-1)^{-1/2}\min_{\bm{G}}\|\bm{G}\widetilde{\bm{C}}^{-1}\bm{D}-\bm{I}\|_F$ over matrices $\bm{G}$ with exactly one non-zero element in each row and column. It is also computed by linear assignment. Rotation matrices are drawn uniformly from the orthogonal group through the QR decomposition of standard Gaussian matrices, with column signs fixed by the diagonal of the triangular factor. Without alignment the two references have the same distribution, since $\|\widetilde{\bm{C}}\bm{R}-\bm{D}\|_F=\|\widetilde{\bm{C}}-\bm{D}\bm{R}'\|_F$ with $\bm{R}'$ again uniformly distributed; the same holds for the minimum-distance index. With alignment they differ: the signed permutation acts on the columns of a rotated ICA estimate in one and on the ICA components in the other. Different FastICA starting values change the ICA estimate by an orthogonal factor, which leaves the first reference unchanged; the second is recomputed for each starting value. The simulated reference generates 20 exactly acyclic systems from the selected coefficient matrix with independent Student-$t_5$ disturbances and 1,990 observations. It then applies the same calculations.

\paragraph{Backward mass.} FastICA estimates an unmixing matrix $W$ on the standardized observations. Rows are matched to variables by a minimum-cost assignment using inverse absolute coefficients, then normalized to a unit diagonal. Let $r(i)$ be a variable's rank in the DirectLiNGAM order. The backward-mass statistic is
\begin{equation}
 R(W,r)=\frac{\sum_{i\ne j}\mathbf{1}\{r(i)<r(j)\}W_{ij}^{2}}
 {\sum_{i\ne j}W_{ij}^{2}}.
\end{equation}
We compare the observed value with the same statistic for 5,000 random orders. Separately, we generate 20 exactly acyclic systems from the selected coefficient matrix with independent Student-$t_5$ disturbances and apply the same ICA alignment procedure. All empirical and simulated ICA fits converge under the recorded numerical tolerance. This reference illustrates finite-sample backward mass even when the generating system is acyclic. Because the empirical order is learned from the same data, the random-order comparison is descriptive rather than a conventional independent hypothesis test. The backward share is 0.399. Across 5,000 random orders, its median is 0.500, and 0.26\% of random orders have a share no greater than the recovered ordering. The corresponding median under 20 simulated exactly acyclic systems is 0.178, with a 5th--95th percentile range of 0.168--0.195. The recovered order is therefore more compatible with the independent-component representation than most random orders, but substantial backward mass remains relative to the simulated reference.

\section{Exact-kernel computation and replication conventions}
\label{app:implementation}
For a centered $n\times n$ kernel matrix $K_c$, define its cyclic diagonals by $D_K(t,k)=K_c(t,(t+k)\bmod n)$. Shifting both indices of a second centered kernel matrix corresponds to shifting the first coordinate of its cyclic diagonals. Consequently, every circular-shift HSIC numerator is a sum of one-dimensional circular cross-correlations:
\begin{equation}
 T(s)(n-1)^2=\sum_k\sum_tD_K(t,k)D_L(t+s,k).
\end{equation}
Fourier transforms evaluate all values of $s$ without approximating the Gaussian kernels. Symmetry of the admissible shift set makes the choice of positive or negative shift convention immaterial to the reference distribution. Unit tests compare the Fourier result with direct Gram-matrix calculations for every shift in a small example.

The replication code sets a global random seed of 1121 and fixes a separate seed for every resampling step: 731 for the coefficient bootstrap, separate outer and inner seeds for each re-discovery replication, and a fixed sequence of alternative orders for the holdout comparator. Results therefore do not depend on the order in which the analyses are run or on the number of parallel workers. The primary coefficient bootstrap has 1,000 replications; re-discovery has 500 outer and 200 inner replications; influence fits have 500 coefficient replications. All empirical block lengths are ten retained observations. The alternative-order comparator uses 1,000 coefficient replications per order, matching the training baseline's budget.

The computational materials include the bank universe and capitalization weights, interval-level observation flags, fitted GARCH parameters, unpruned and selected coefficient matrices, pairwise exact-HSIC outputs, re-discovery ranks, and all sensitivity outputs. The implementation is stated explicitly so that software defaults for coefficient shrinkage, feature approximations, and variance scaling do not silently change the estimand. Regenerating the empirical pipeline requires the same underlying price vintage and sufficient scratch space for exact kernel calculations.

\end{document}